\documentclass{ws-mpla}
\usepackage[super]{cite}
\usepackage{graphicx}
\usepackage{hyperref}
\usepackage{color}
\usepackage{lineno}

\begin{document}

\markboth{Andrea Dubla, Umut G{\"u}rsoy, Raimond Snellings}
{Charge-dependent flow as evidence of strong electromagnetic fields in heavy-ion collisions}

\catchline{}{}{}{}{}

\title{CHARGE-DEPENDENT FLOW AS EVIDENCE OF STRONG ELECTROMAGNETIC FIELDS IN HEAVY-ION COLLISIONS}

\author{\footnotesize ANDREA DUBLA}

\address{GSI Helmholtzzentrum f{\"u}r Schwerionenforschung, 64291 Darmstadt, Germany\\
Nikhef National Institute for Subatomic Physics,  Sciencepark 105, 1098 XG, Amsterdam, The Netherlands\\
andrea.dubla@cern.ch}

\author{UMUT G{\"U}RSOY}
\address{Institute for Theoretical Physics, Utrecht University, Leuvenlaan 4, 3584 CE Utrecht, The Netherlands\\
u.gursoy@uu.nl 
}

\author{RAIMOND SNELLINGS}
\address{Institute for Gravitational and Subatomic Physics (GRASP),
Utrecht University, Princetonplein 1, 3584 CC Utrecht, The Netherlands\\
Nikhef National Institute for Subatomic Physics,  Sciencepark 105, 1098 XG, Amsterdam, The Netherlands\\
raimond.snellings@cern.ch 
}

\maketitle

\pub{Received ( 24 September 2020)}{Published (13 October 2020)}

\vspace{5 mm}

\begin{abstract}
The extremely large electromagnetic fields generated in heavy-ion collisions provide access to novel observables that are expected to constrain various key transport properties of the quark-gluon plasma and could help solve one of the outstanding puzzles in QCD: the strong CP problem. 
In this review we present a brief overview of the theoretical  and experimental characterization of these electromagnetic fields. 
After reviewing the current state, emphasising one of the observables -- the charge-dependent flow -- we discuss the various discrepancies between the measurements and theoretical predictions. Finally, to help resolve the discrepancies, we suggest new measurements and theoretical ideas. 

\end{abstract}

\ccode{PACS Nos.: 25.75.-q, 25.75.Nq}

\newpage

\section{Introduction}	

Quantum chromodynamics (QCD) calculations on the lattice \cite{Bazavov:2014pvz, pQCD2, pQCD3} predict at high temperature and energy density the existence of a deconfined state of quarks and gluons, known as the quark-gluon plasma (QGP). Characterizing thermodynamic and transport properties of the QGP is among the main goals of the ultra-relativistic heavy-ion experimental program at the Relativistic Heavy Ion Collider (RHIC) and the Large Hadron Collider (LHC).
Ultra-relativistic heavy-ion collisions with a non-zero impact parameter are characterized by extremely strong electromagnetic fields primarily induced by spectator protons\footnote{Spectator protons are protons from the incoming nuclei which do not undergo inelastic collisions}.
Currently, there is strong interest in characterizing the time evolution of these fields which are estimated, via application of the Biot-Savart law to heavy-ion collisions, to reach up to $\rm 10^{18}$ -- $\rm 10^{19}$ Gauss \cite{Tuchin:2013ie,Skokov:2009qp}. 
Several extremely interesting quantum field theoretical phenomena are predicted to occur in the presence of such strong electromagnetic fields, including the chiral magnetic effect (CME), that is the generation of an electric current along the magnetic field in a medium with chiral imbalance~\cite{Fukushima:2008xe,Kharzeev:2007jp,Tuchin:2010vs,Voronyuk:2011jd} and chiral magnetic wave (CMW), that is a collective gapless excitation arising from the coupling between the density waves of the electric and chiral charges~\cite{Kharzeev:2010gd}. 

Observables constructed to be sensitive to these phenomena, such as the induced charge dependent correlations~\cite{Voloshin_2004} due to CME, have been measured~\cite{Abelev:2009ac,Adamczyk:2014mzf,Abelev:2012pa, Acharya:2017fau, Khachatryan:2016got,Sirunyan:2017quh} and are in qualitative agreement with theoretical expectations~\cite{Toneev:2010xt}. However, the limited precision of the measurements and, in addition, the possible background contributions, prohibit unequivocal claims that the origin of these observed charge dependent correlations is due to CME.
To make progress it is important to establish other observable consequences of the early-time magnetic field on the final-state charged particles that are preferably due to alternative transport phenomena and which would allow us to calibrate the strength and lifetime of the magnetic field independently.
One way to disentangle the measurement of the electromagnetic field from the exotic effects, is to look at the electromagnetic currents that this field would induce in the QGP constituents. 
Such electromagnetically induced currents are expected to be present quite generically -- unlike CME or CMW which require in addition a chiral imbalance -- as they originate from common physics laws such as the Lorentz force\footnote{ that is the force exerted by the external and internal electromagnetic fields on the charged particles floating in the plasma}.  

Our paper is organised as follows: In section \ref{theory} we give for light- and heavy-flavor particles an overview of the theory and model calculations of the observables sensitive to the electromagnetic field. In section \ref{exper} we report a summary of the measurements at RHIC and LHC energies, together with their possible interpretation. 
Finally, in section \ref{future}, we discuss new ideas and measurements, which have the potential to resolve the current mismatch between the measurements and the theoretical calculation.

\section{Theory model calculations and predictions}
\label{theory}

To constrain the strength of the electromagnetic fields experimentally one should first assess their observable effects theoretically. It was first shown in Ref.~\refcite{Gursoy:2014aka} that the magnetic field produced in a heavy-ion collision could result in a measurable effect in the form of a charge-odd contribution to the directed flow coefficient $\Delta v_1$. The directed flow $v_1$ itself is the first coefficient of the Fourier expansion of the azimuthal distribution of produced particles, and is odd in rapidity for symmetric collisions. The electromagnetically induced component of this flow, $\Delta v_1 = v_1^+ - v_1^-$, that is the difference between the positively and negatively charged particles of the same mass e.g. $\pi^+$ and $\pi^-$, is constructed to be independent of the background flow and hence should be directly sensitive to the electromagnetically induced transverse currents in the plasma.
In Ref. \refcite{Gursoy:2018yai} this charge-odd contribution was studied via $\Delta v_1$ and the other higher harmonic charge-odd flow coefficients $\Delta v_n$. 

\begin{figure}[t]
\vspace{-0.2in}
 \begin{center}
\includegraphics[scale=0.45]{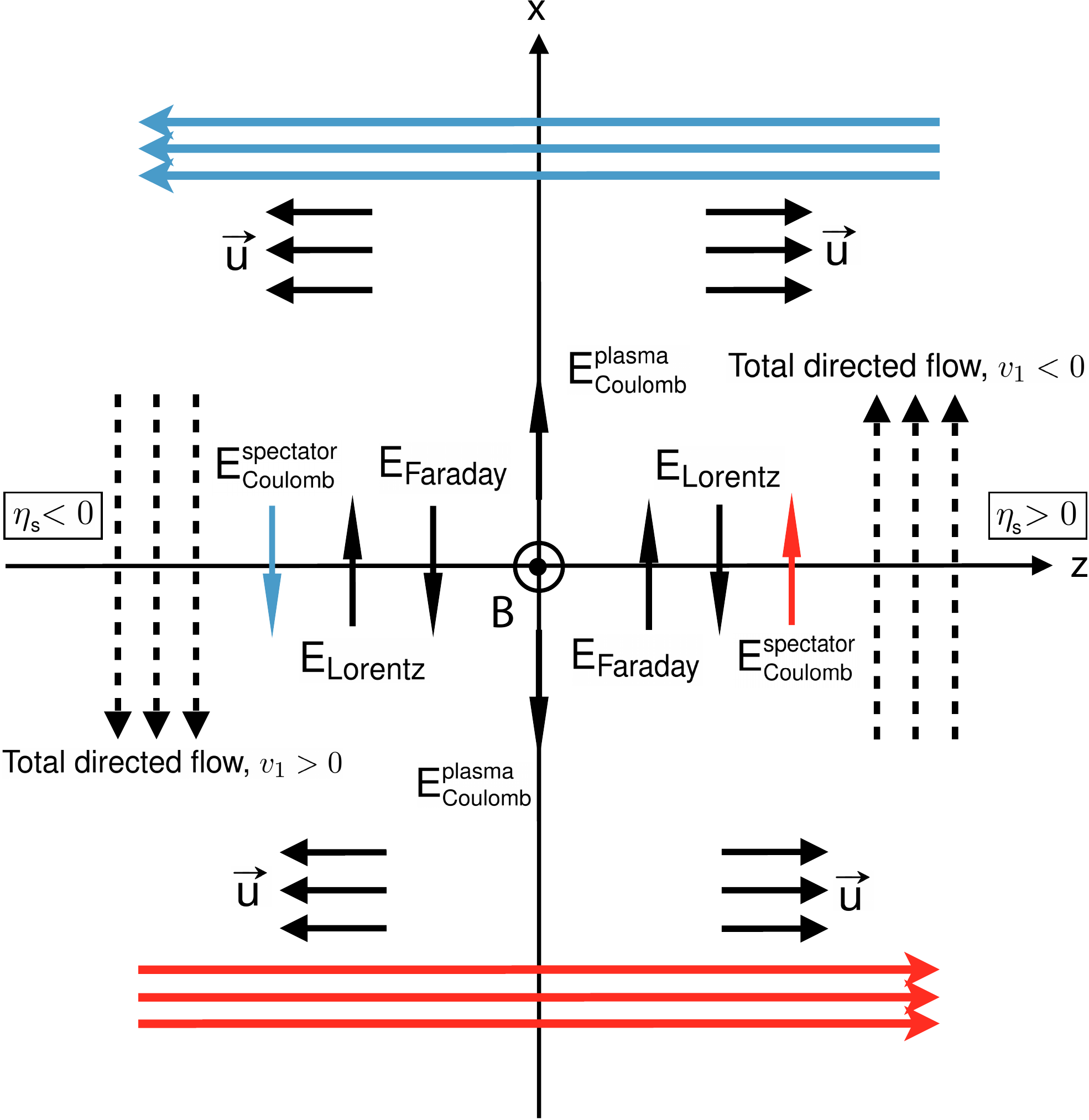}
 \end{center}
 \vspace{-0.2in}
 \caption{Schematic illustration of how the magnetic field $\vec B$ in a heavy-ion collision results
in a directed flow {\it of electric charge}, $\Delta v_1$.  The collision occurs in the $z$-direction, resulting in a longitudinal expansion velocity $\vec u$ of the QGP produced in $+z$ ($-z$) the direction at positive (negative) $z$.  We take the impact parameter vector to point in the $+x$ direction,
choosing the nucleus moving toward positive (negative) $z$ to be located at  negative (positive) $x$. 
The trajectories of the spectators that ``miss'' the collision because of the nonzero impact parameter are indicated by the red and blue arrows.
This configuration generates a magnetic field $\vec B$ in the $+y$ direction, as shown.
The directions of the electric fields (and hence currents)
due to the Faraday, Lorentz and Coulomb effects are shown.}
\label{figschema}
\end{figure}
This charge-odd $\Delta v_n$ arises from four different contributions that are schematically described in Fig.~\ref{figschema}:
i) {\it Faraday current} which arises from the decrease of the magnetic field with time, dictating induction of an electric current by Faraday's law; (ii) {\it Lorentz current} which arises from the longitudinal flow velocity $\vec{v}_\mathrm{flow}$ (denoted by $\vec u$ in Fig.~\ref{figschema}) of the hydrodynamic fluid and the Lorentz force; (iii) {\it Coulomb current} arising from the positively charged spectators that have passed the collision zone exerting an electric force on the charged plasma produced in the collision; and finally (iv) {\it Plasma current}, an outward component of the electric field that originates from the net positive charge in the plasma. As explained in Fig.~\ref{figschema} the first three effects combined are expected to result in an electromagnetically induced charge-odd directed flow, $\Delta v_1$, whereas the last effect results in a non-trivial electromagnetically induced radial and elliptic flow,  $\Delta \langle p_T \rangle$ and $\Delta v_2$. 

The magnitude of these effects can be estimated by following a perturbative approach to relativistic magneto-hydrodyamics \cite{Gursoy:2014aka}, i.e. by appending the motion of charges induced by the electromagnetic fields on the hydrodynamic background that is calculated in the absence of the electromagnetic fields. With the additional assumption that the electromagnetic interactions can be treated classically \footnote{This is controlled by the total magnetic energy of the medium compared to the energy of a single photon with wavelength comparable to the size of the medium. The ratio can be shown to vary between ~1000 to ~50 as the proper time $\tau$ increases from 0.3 fm to 0.8 fm.}, the electromagnetic fields after the collision are obtained by solving the Maxwell's equations in the conducting medium sourced by the spectator and the participant nucleons. One important ingredient is the electric conductivity of the QGP, which according to lattice QCD calculations has a mild dependence on the temperature  \cite{Ding:2010ga,Francis:2011bt,Brandt:2012jc,Amato:2013naa,Buividovich:2020dks,Aarts:2020dda}. 
As shown in Ref. \refcite{Tuchin:2013ie} (see also Ref. \refcite{Gursoy:2014aka}) a nontrivial electric conductivity in the medium delays the decay of the magnetic field significantly, strengthening magnetically induced phenomena. Therefore, the magnetically induced flow harmonics can, in principle, be used to constrain the value of the QGP conductivity.

Maxwell's equations for a single proton, moving in the $z$-direction with velocity $v$ and located on the interaction plane at $\vec x'_\perp$, read~\footnote{In Ref.~\refcite{Gursoy:2014aka} the electric conductivity $\sigma$ was estimated by the fixed value of $0.023~\mathrm{fm}^{-1}$ corresponding to $\sigma$ at $T=255$ MeV found in lattice studies~\cite{Ding:2010ga,Francis:2011bt,Brandt:2012jc,Amato:2013naa}.}
\begin{eqnarray}
\label{max1} 
{} & & \nabla\cdot \vec{B} = 0,  \qquad \nabla \times \vec{E} = - \frac{\partial\vec{B}}{\partial t}\,  \\
\label{max2} 
{} & & \nabla\cdot \vec{E} = e \delta(z-vt)\delta(\vec x_\perp-\vec x'_\perp) \, \\ 
\label{max3} 
{} & & \nabla \times \vec{B} = \frac{\partial\vec{E}}{\partial t} +  \sigma  \vec{E} + ev \hat{z} \delta(z-vt)\delta(\vec x_\perp-\vec x'_\perp)  \, .
\end{eqnarray}
These equations can be solved analytically using the Green's functions \cite{Tuchin:2010vs, Tuchin:2013ie, Gursoy:2014aka} and results in the magnetic field arising from a single proton (moving in the +$z$ direction)
\begin{equation}\label{Bfull}
 e B^+_y = \frac{\alpha_{em}\gamma^4v^4}{8}|x_\perp - x'_\perp|\cos\alpha \frac{(\sigma\sqrt{U^2+V^2} +1)}{(U^2+V^2)^{\frac32}} e^{\sigma\left(U-\sqrt{U^2+V^2}\right)}\, ,
 \end{equation}
where 
\begin{equation}\label{defUV}
U \equiv \frac{\gamma^2 v^2}{2}(t-\frac{z}{v})\, , \qquad V\equiv \frac{\gamma v}{2}|\vec x_\perp-\vec x'_\perp|\, ,
\end{equation}
and $\alpha$ is the angle between $x_\perp - x'_\perp$ and the $x$-axis. The $z$ component of the magnetic field vanishes and the $x$ component is obtained by replacing $\cos(\alpha)$ with $-\sin(\alpha)$ in the expression above. Similar analytic expressions for the electric fields sourced by a single proton can be found in Ref.~\refcite{Gursoy:2014aka}. The total electromagnetic fields are then found by integrating over the spectator and the participant proton distributions. In Ref.~\refcite{Gursoy:2018yai} these distribution profiles are generated for $10^{4}$ events using the Monte-Carlo Glauber model
~\cite{Shen:2014vra} that is also used to initialize the hydrodynamic calculation. The spectators are assumed to move with the beam rapidity $Y = \tanh^{-1} v$, whereas the rapidity loss of the participant nucleons is assumed to follow the phenomenological distribution \cite{Kharzeev:2007jp,Fukushima:2008xe}
\begin{equation}
f^{\pm}(y) = \frac{1}{4 \sinh(Y/2)} e^{\pm y/2} \quad \mbox{for} \quad  -Y< y <Y\, .
\label{ParticipantRapidityDistribution}
\end{equation}

\begin{figure}[t!]
 \begin{center}
\includegraphics[scale=0.3]{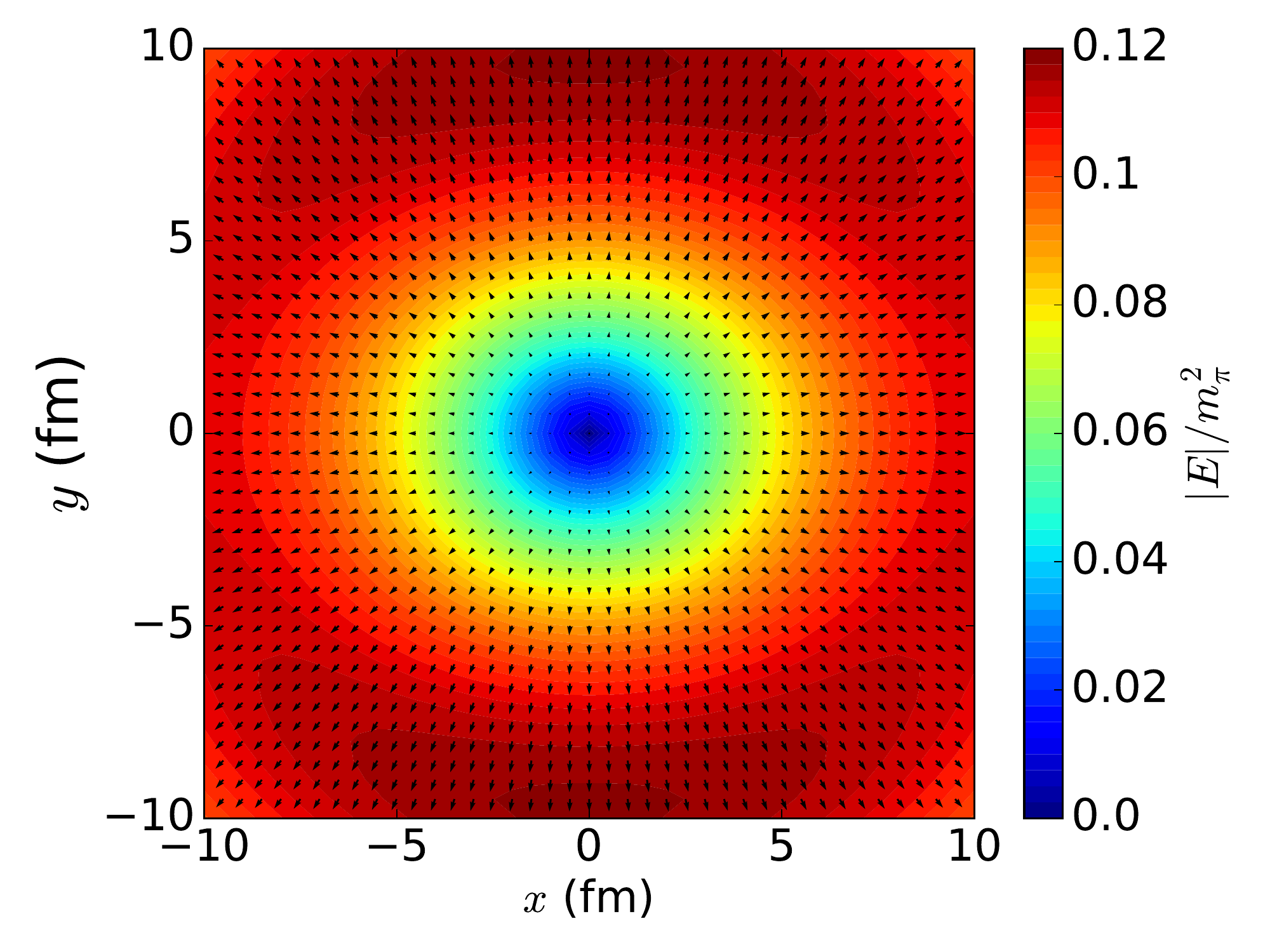}
\includegraphics[scale=0.3]{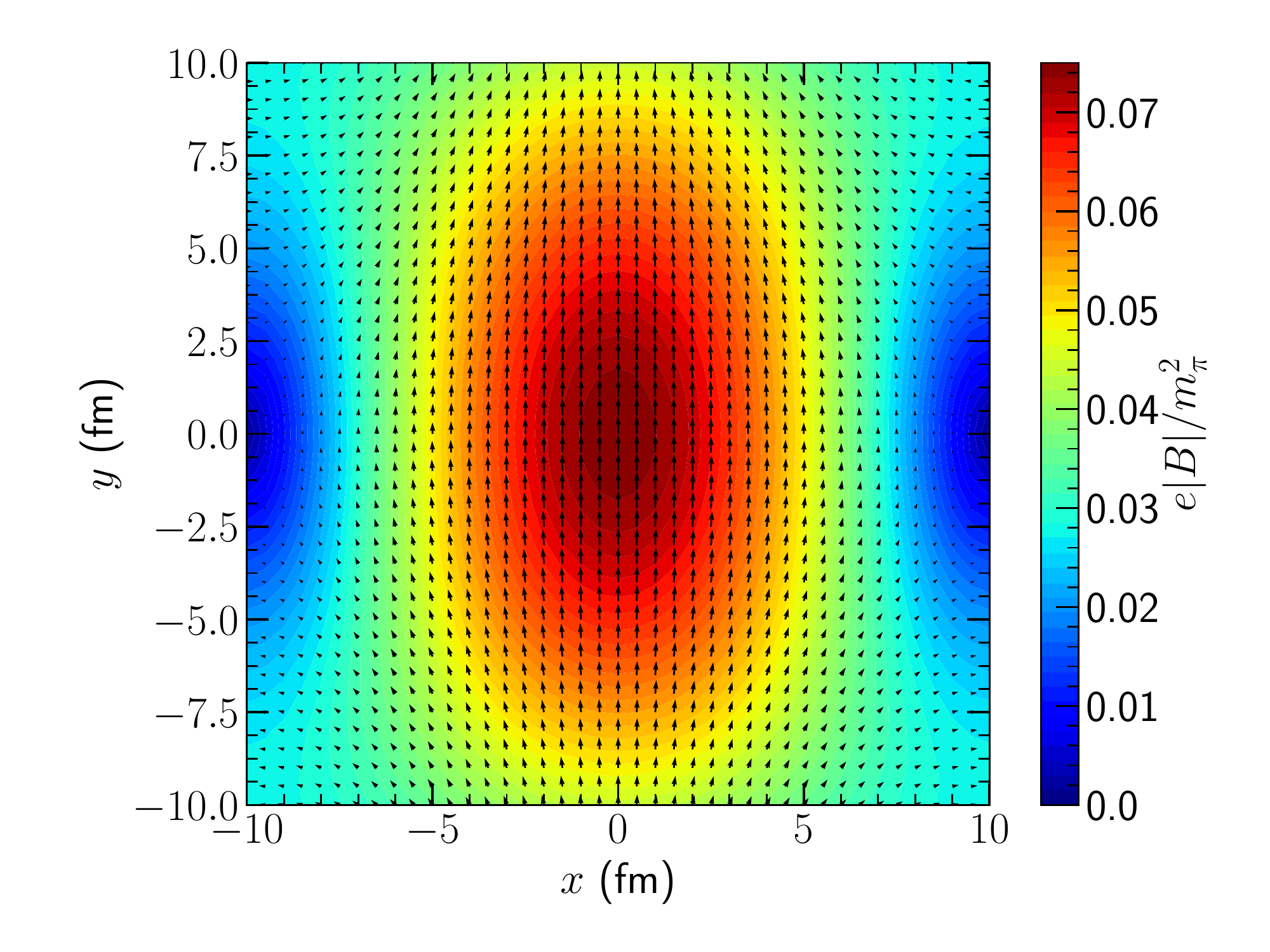}
 \end{center}
 \caption[]{The electric (left) and magnetic (right) fields in the transverse plane at $z=0$  in the lab frame 
 at a proper time $\tau=1$~fm/c after a Pb--Pb collision
 with 20--30\% centrality (corresponding to impact parameters in the range 6.24~fm~$<b<$~9.05~fm) and with a collision energy $\sqrt{s}$ = 2.76 TeV.}
\label{figEB}
\end{figure}

\begin{figure}[t!]
 \begin{center}
\includegraphics[scale=0.3]{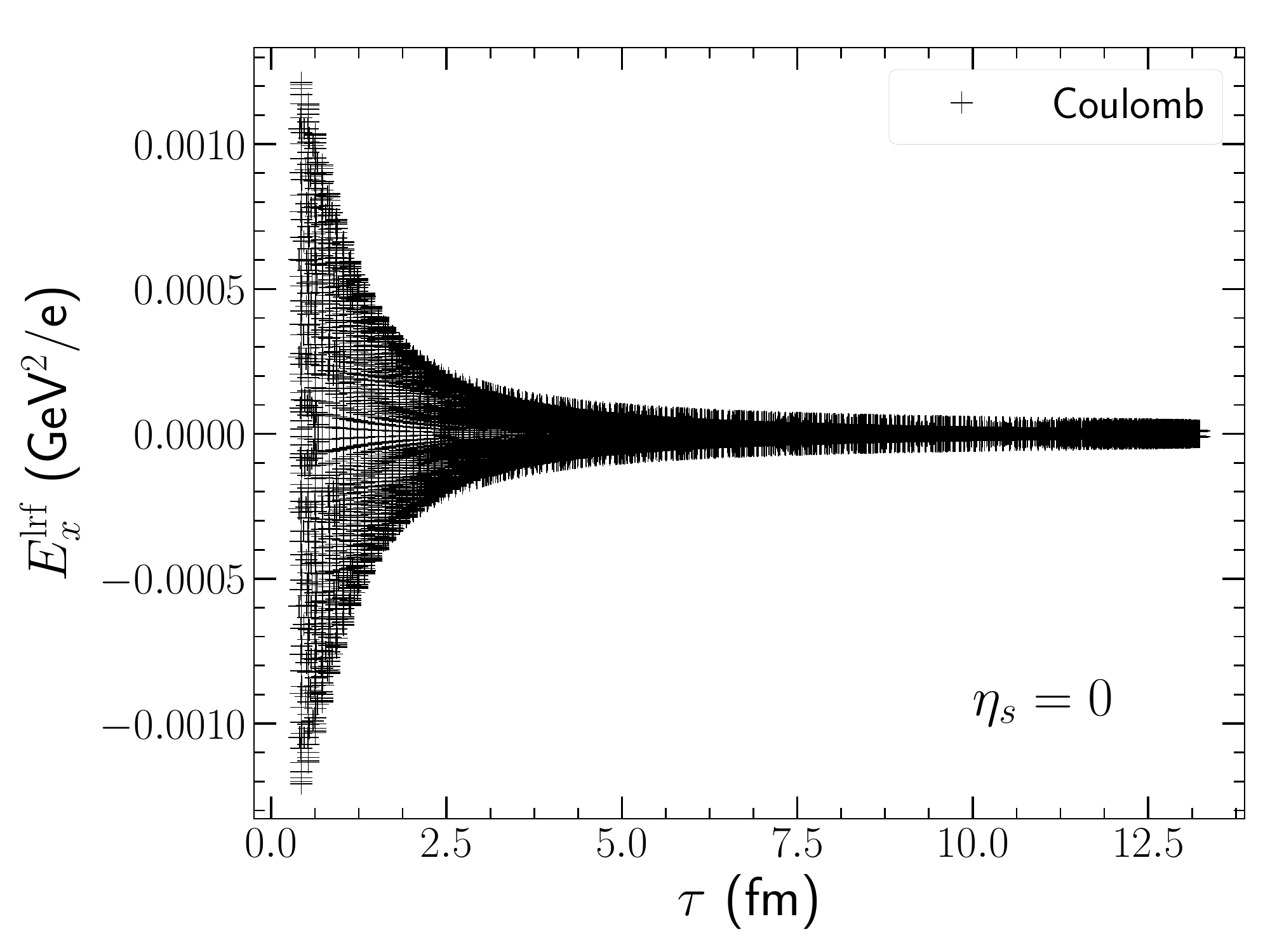}
\includegraphics[scale=0.3]{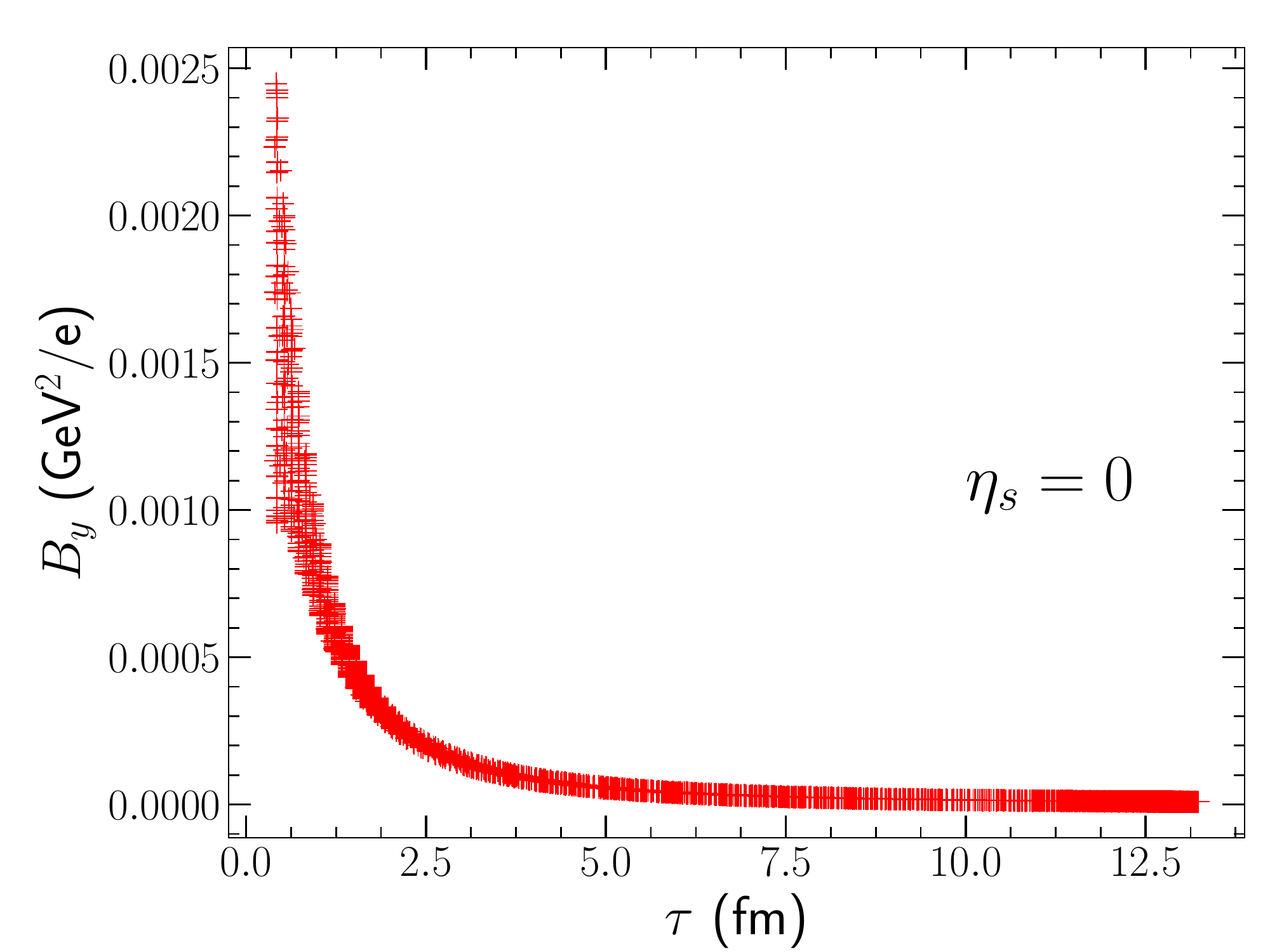}
 \end{center}
 \caption[]{Time dependence of the $y$-component of the electric (left) and magnetic field (right) in the lab frame at $\eta_s=0$. Each cross corresponds to a single point on the freeze-out surface.} 
\label{figEBt}
\end{figure}

The resulting profile of the electromagnetic fields in the transverse plane is shown in Fig.~\ref{figEB}.  The fields are produced by the spectator protons moving in the $+z$ ($-z$) direction for $x<0$ ($x>0$) as well as by the ones that participate in the collision. The direction of the fields are shown by the black arrows and their strengths are indicated both by the length of the arrows and by the colour.  The magnetic field is strongest at the center of the plasma, where it points in the $+y$ direction as was shown in Fig.~\ref{figschema}. The electric field points mostly outward and is strongest on the periphery of the plasma.
The time profile of the magnetic field at vanishing space-time rapidity is plotted in Fig.~\ref{figEBt}, where each cross represents a point on the freezout surface of the expanding fluid.

In the perturbative scheme for magneto-hydrodynamics that we follow here, the background fluid can be obtained from hydrodynamics in the absence of electromagnetic fields. In Ref. \refcite{Gursoy:2018yai} 
the dynamical evolution of the medium was obtained using the iEBE-VISHNU framework \cite{Shen:2014vra}. Taking event-averaged initial conditions from a Monte-Carlo Glauber model and averaging
over $10^{4}$ events one generates viscous hydrodynamic solutions which are in remarkable agreement with data from measurements at the highest RHIC and LHC energies. See Ref. \refcite{Gursoy:2018yai} for the details. 

The additional velocity component $\vec v_{\rm EM}$ of the fluid which arises from the electromagnetic effects can be added to the background velocity either by solving a force-balance law \cite{Gursoy:2014aka} or by evoking Langevin dynamics \cite{Das:2016cwd, Chatterjee:2018lsx} for the light- or the heavy-flavor particles, respectively. Below, we explain the model calculations in each case and present the corresponding results.


Assuming the electromagnetically induced velocity is only a small perturbation compared to the background hydrodynamic flow, $\vert \vec{v}_\mathrm{EM} \vert \ll \vert \vec{v}_\mathrm{flow} \vert$, we can obtain $\vec v_\mathrm {EM}$ for the light quarks by solving the force-balance equation~\cite{Gursoy:2014aka} 
\begin{equation}
m \frac{d \vec{v}_\mathrm{EM}}{dt} = q \vec{v}_\mathrm{EM} \times \vec{B}^\mathrm{\, lrf} + q \vec{E}^\mathrm{\, lrf} - \mu m \vec{v}_\mathrm{EM} = 0
\label{eq9}
\end{equation}
in its non-relativistic form at any space-time point in the local rest frame of the fluid cell. Here the superscript $\mathrm{lrf}$ refers to "local rest frame". This equation balances the Lorentz force with  the drag in a fluid element with mass $m$. The value of the drag coefficient $\mu$ is fixed in Refs.~\refcite{Gursoy:2014aka,Gursoy:2018yai} using the analogous strong coupling  value in ${\cal N}=4$ supersymmetric Yang-Mills theory \cite{Herzog:2006gh,CasalderreySolana:2006rq,Gubser:2006bz}. 
Finally, the EM velocity $\vec{v}_\mathrm{EM}$ in every fluid cell along the freeze-out surface  is boosted by the flow velocity to bring it back  to the lab frame, which is then used as the input in the standard Cooper-Frye procedure \cite{Cooper:1974mv} to calculate the hadron spectra and the flow coefficients. We present a selection of results in comparison to experimental data in the next section. 


As first argued in Ref.~\refcite{Das:2016cwd} electromagnetically induced flow on the heavy quarks is expected to be several orders of magnitude larger than that for the light quarks because of (i) earlier formation time when the electromagnetic fields are stronger and (ii) longer equilibration time which allows them to keep the initial kick they receive from the electromagnetic force at the time of freezout. Therefore, charm quarks, who are the most abundant heavy-flavor quarks and with charge $q$~=~2/3, provide crucial and independent information on the strength of the magnetic field and on the distribution of the matter produced in heavy-ion collisions. To study electromagnetically induced directed flow for the heavy-flavor quarks  Langevin dynamics~\cite{Rapp:2009my, Moore:2004tg}, instead of the force-balance law, is more suitable because the stochastic forces cannot be ignored now. The Langevin process updates the position $\Delta r$ and momenta $p$ of the heavy-flavor quarks with energy $E$ at any given time $t$ as 
\begin{eqnarray}\label{Langevin} 
\Delta r_i &=& \frac{p_i}{E} \Delta t \\
\Delta p_i &=& - \gamma p_i \Delta t +  \rho_i \sqrt{2D\Delta t} +  F^\mathrm{EM}_i
\end{eqnarray}
where the stochastic force coefficients satisfy $\langle \rho_i \rangle =0$ and   $\langle \rho_i \rho_j \rangle = \delta_{ij}$. $F^\mathrm{EM}$ denotes the electromagnetic force  in the local reference frame of the quark. Finally $\gamma$ and $D$ are the drag and the diffusion coefficients that satisfy the Einstein relation 
\begin{equation}\label{Einstein} 
D = \gamma E T\, .
\end{equation}
The results of this procedure are presented in the next section where we also compare them with experimental measurements.  

Among further work which we do not cover here, due to restricted space, is the recent investigation of small and asymmetric collisions, such as proton--gold, that were treated by the Parton-Hadron-String Dynamics (PHSD) transport approach~\cite{Oliva:2019kin,Oliva:2020mfr}, where the parton dynamics follows from the Dynamical QuasiParticle Model (DQPM)~\cite{Cassing:2007nb}. These are microscopic off-shell transport models based on a quasi-particle description of the plasma. As such, they differ from the hydrodynamics approach reviewed in this section.

\section{Experimental measurements and theory comparison}
\label{exper}

To measure the directed flow two different techniques have been used, the STAR Collaboration at RHIC used the event plane method while the ALICE Collaboration at the LHC computed the directed flow using the scalar-product method~\cite{Luzum:2012da}.
For both experimental techniques, two zero degree calorimeters, one located at forward and the other at backward pseudorapidy, are used to measure the transverse distribution of the spectator neutrons as well as the energy they deposited. The direction of deflection of the spectator neutrons is estimated event-by-event with the flow vectors, ${\rm\bf Q}^{t,p}$, where {\it p (t)} denotes the projectile (target) side~\cite{Abelev:2013cva}. The directed flow is then calculated as: $v_{1} = ( v_1^{ p} - v_1^{ t}$)/2.
At both RHIC and the LHC, the established convention is that a positive sign of $v_1$ is defined relative to the deflection of the projectile spectators. The measurement of $v_1$ using spectators does not require any treatment of momentum conservation, unlike the measurements based on correlations between particles produced at midrapidity~\cite{Retinskaya:2012ky}.

\subsection{Light-flavor measurements}

The STAR Collaboration measured charge-dependent directed flow for positive and negative charged pions and for protons and antiprotons \cite{Adamczyk:2014ipa,Adamczyk:2017nxg}. 
\begin{figure}[ht!]
\centering
\includegraphics[scale=0.37]{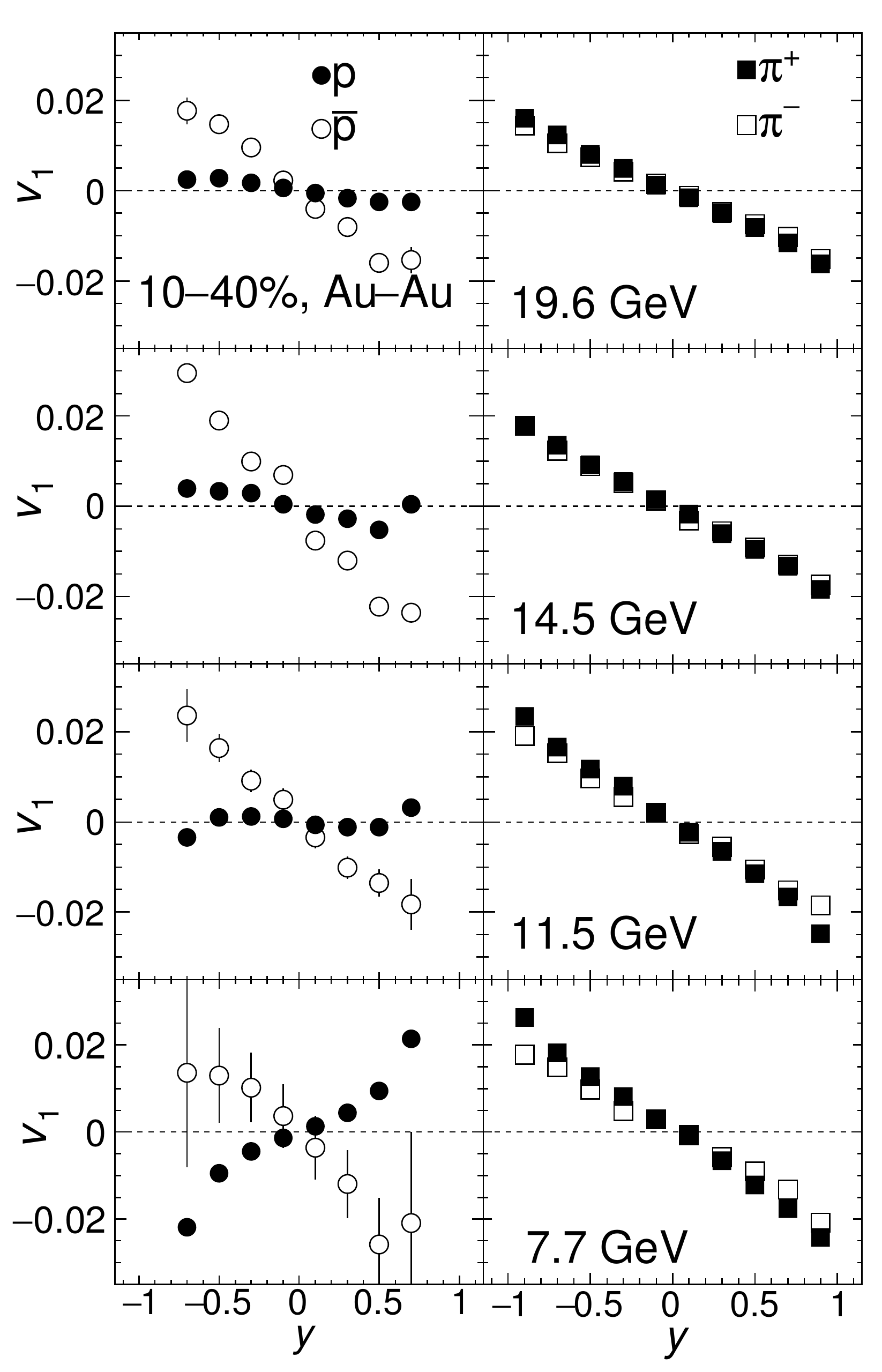}
\includegraphics[scale=0.37]{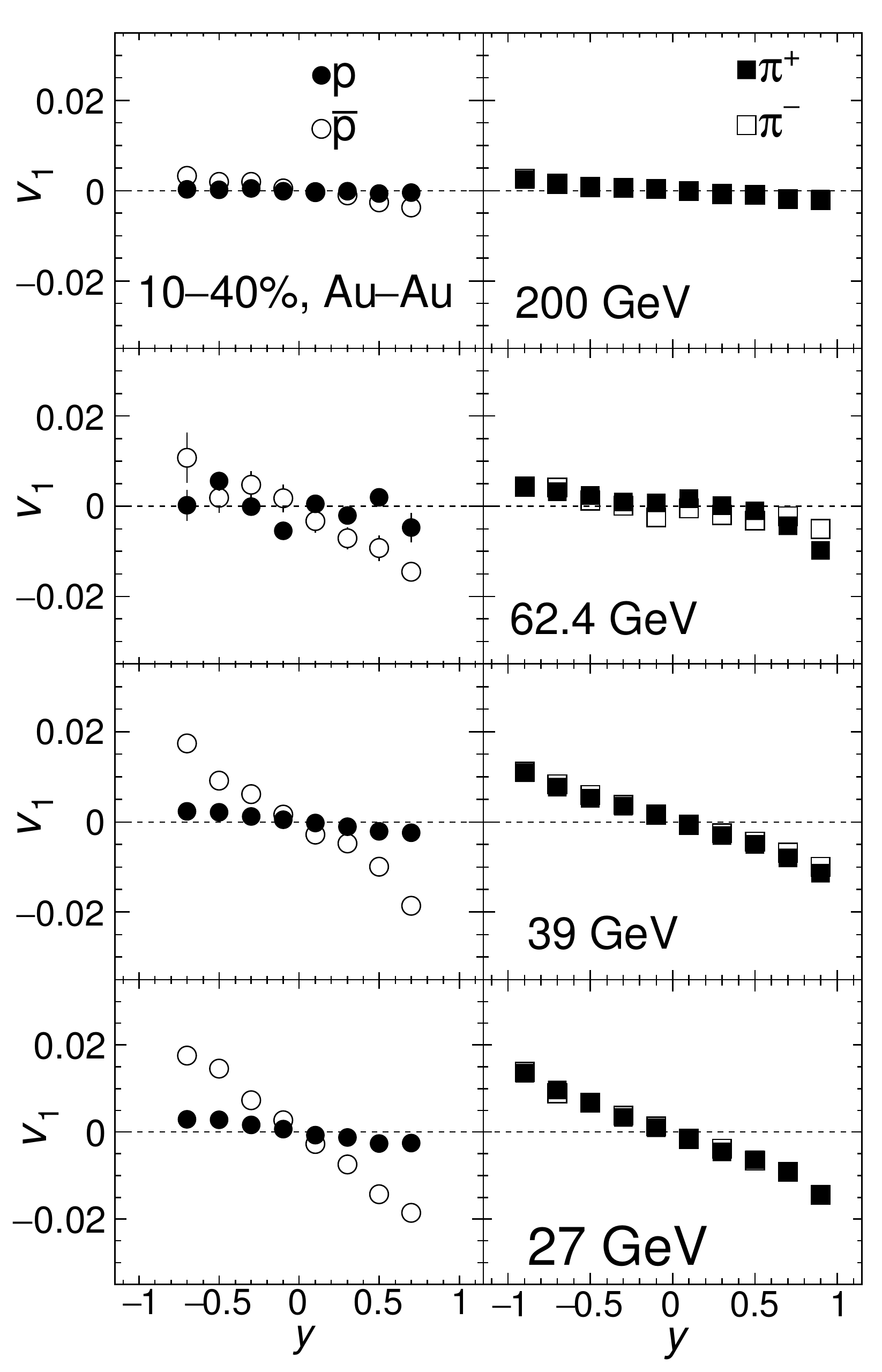}
\caption{Proton and antiproton $v_1$ and
$\pi^{\pm}$ $v_1$ as a function of rapidity for the 10--40\% centrality interval in 
Au--Au collisions at 200, 62.4, 39, 27, 19.6, 14.5, 11.5 and 7.7 GeV.}
\label{starv1pionproton}
\end{figure}
In Fig. \ref{starv1pionproton} the $v_1$ as a function of rapidity for protons (p), antiprotons ($\bar{\rm p}$) and pions ($\pi^{\pm}$) is
plotted for the 10--40\% centrality interval in Au--Au collisions for collision energies ranging from 7.7 GeV up to the top RHIC energies of 200 GeV per nucleon pair. 
A significant difference between proton and antiproton $v_1$ at all eight collision energies is found. It is very interesting to notice that the antiproton $v_1$ has a negative slope at all energies, while for the proton $v_1$ a changes in sign is observed between 7.7 and 11.5~GeV. For positively and negatively charged pions the $v_1$ coefficients have a negative slope as function of rapidity ($y$) and are very close at high energies, with small differences at 11.5 and 7.7~GeV. 

\begin{figure}[ht!]
\centering
\includegraphics[scale=0.3]{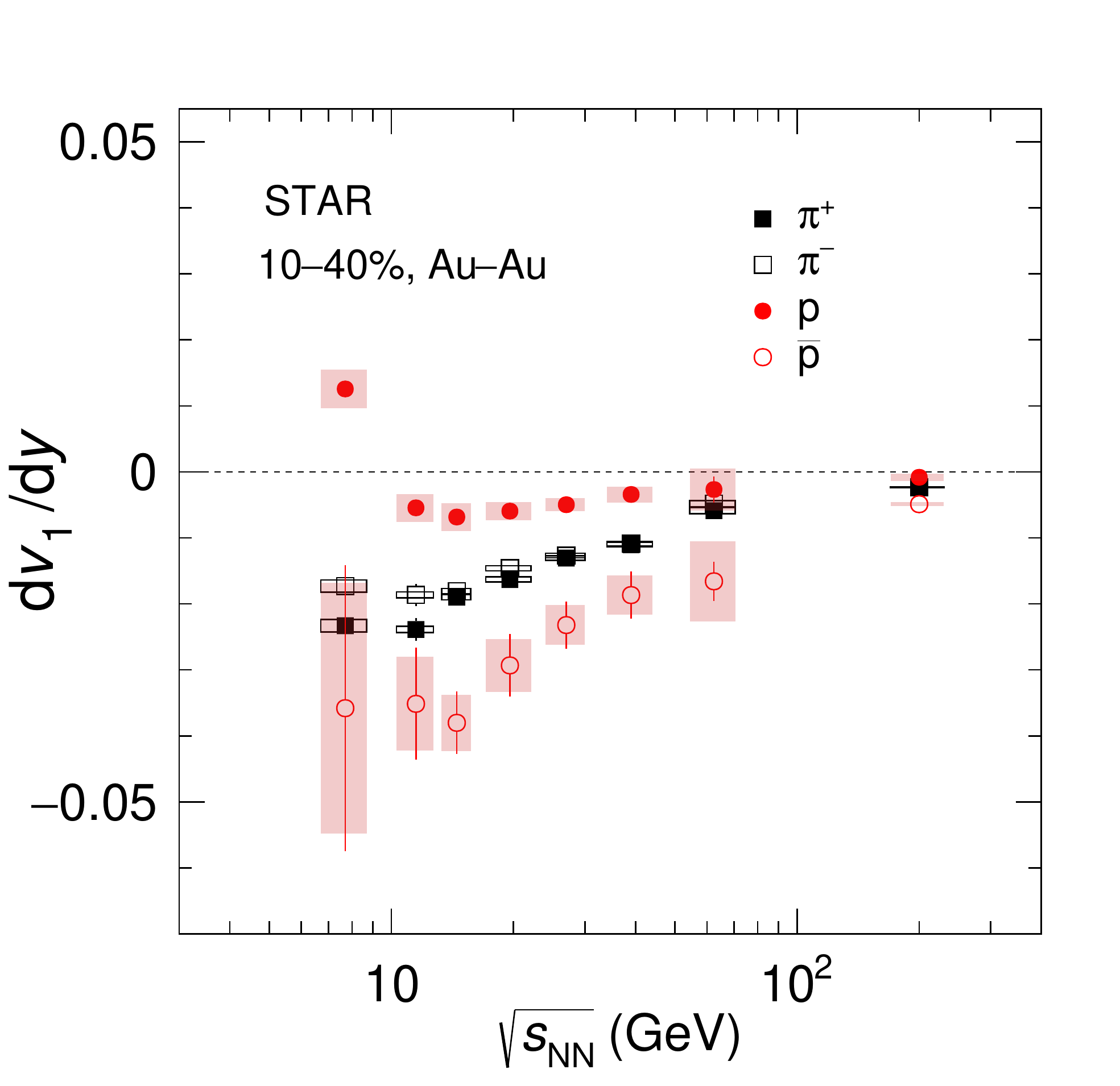}
\includegraphics[scale=0.3]{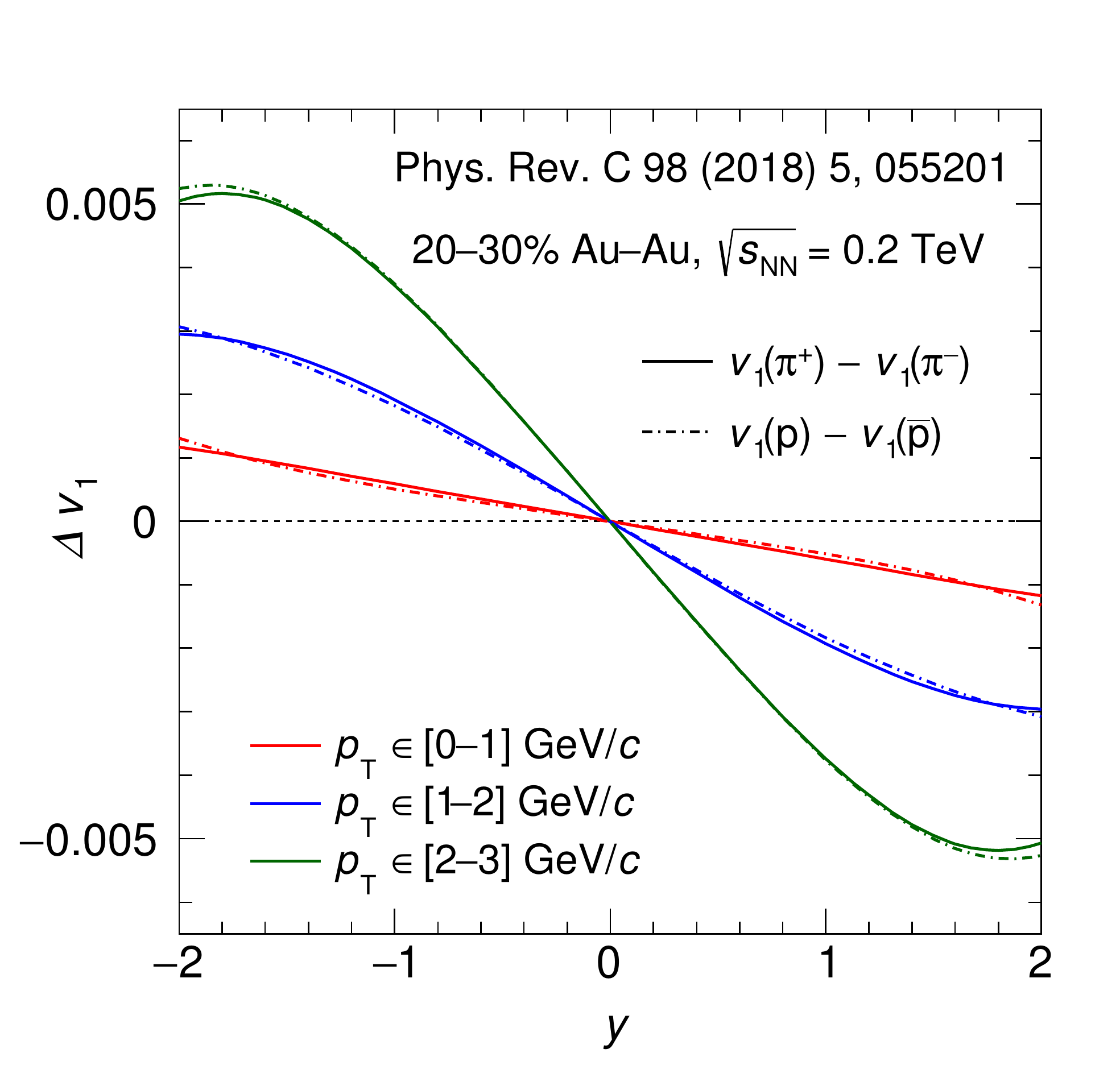}
\caption{ (Left panel) Directed flow slope (d$v_1$/d$y$) for charged pions, protons and antiprotons near midrapidity as a function of the beam energy for the 10--40\% centrality interval in Au--Au collisions.
(Right Panel) The electromagnetically induced difference between $v_1$ coefficients of $\pi^+$ and $\pi^-$ mesons (solid lines) and between protons and antiprotons (dash-dotted lines) as a function of particle rapidity for 20--30\% Au--Au collisions at $\sqrt{s_{\rm NN}}$ = 200 GeV. Three different $p_{\rm T}$ integration ranges are shown for each of the $\Delta v_1$.}
\label{starv1modelcomp}
\end{figure}
To extract the $v_1$ slope, d$v_1$/d$y$, and to better visualise its energy dependence, the $v_1$ of protons, antiprotons, and $\pi^{\pm}$ was fit with a cubic function, where the coefficient of the linear term represents the slope. The fit was performed between $-$0.5$~< \eta~<$~0.5.
The slopes for protons, antiprotons and $\pi^{\pm}$ are shown in the left panel of Fig. \ref{starv1modelcomp} as a function of the collision energy.
The figure clearly shows how the difference in the $v_1$ between proton and antiproton decreases with increasing energy, as well as a change in slope for protons between the collision energies 7.7 and 11.5 GeV. The slope of $v_1$ for protons shows a minimum between 11.5 and 19.6 GeV and remains small and negative up to 200 GeV. The difference between the positive and negative charged pions also decrease with energy, this difference is already small at the lowest energy and disappears within uncertainties at~19.6~GeV.

The pion $\Delta v_1$ from the model calculations for Au--Au at $\sqrt{s_{\rm NN}}$~=~200~GeV, shown in the right panel of Fig.~\ref{starv1modelcomp}, shows a negative slope as a function of rapidity, consistent with the experimental findings from STAR. 
A more detailed comparison for the pions at RHIC currently cannot be made because of low statistical significance of the measurements at $\sqrt{s_{\rm NN}}$~=~200~GeV. 
The same model calculation for protons, also shown in the right panel of Fig.~\ref{starv1modelcomp}, disagrees with the STAR measurement in the sign. 
The predicted sign for the $\Delta v_1$ of protons is negative whereas the measured experimental value of $v_1({\rm p})$ minus $v_1(\rm{\bar p})$ is positive. 
Nevertheless, this disagreement currently does not rule out the contribution from the EM fields as calculated in Ref.~\refcite{Gursoy:2018yai}. This is because the proton d$v_1$/d$y$ and its energy dependence is currently poorly understood. The proton $v_1$ receives contributions from baryons transported from the beam rapidity to the vicinity of midrapidity and from the protons from particle-antiparticle pair production. Clearly the importance of the second mechanism increases strongly with increasing beam energy~\cite{Adamczyk:2011aa,Adamczyk:2014ipa,Adamczyk:2016eux}. Unfortunately, even with important recent net-proton $v_1$ measurements from Ref. \refcite{Adamczyk:2014ipa}, the contribution of these two mechanisms to $v_1$ is currently still not completely understood. 
To better understand the possible role and relevance of baryon number transport, $v_1$ measurements as a function of centrality and in a larger rapidity range, together with a systematic comparison to hydrodynamical model calculations which include baryon number transport~\cite{Shen:2020jwv,Li:2018ini,McLerran:2018avb} would be very important.

\begin{figure}[ht!]
\centering
\includegraphics[scale=0.3]{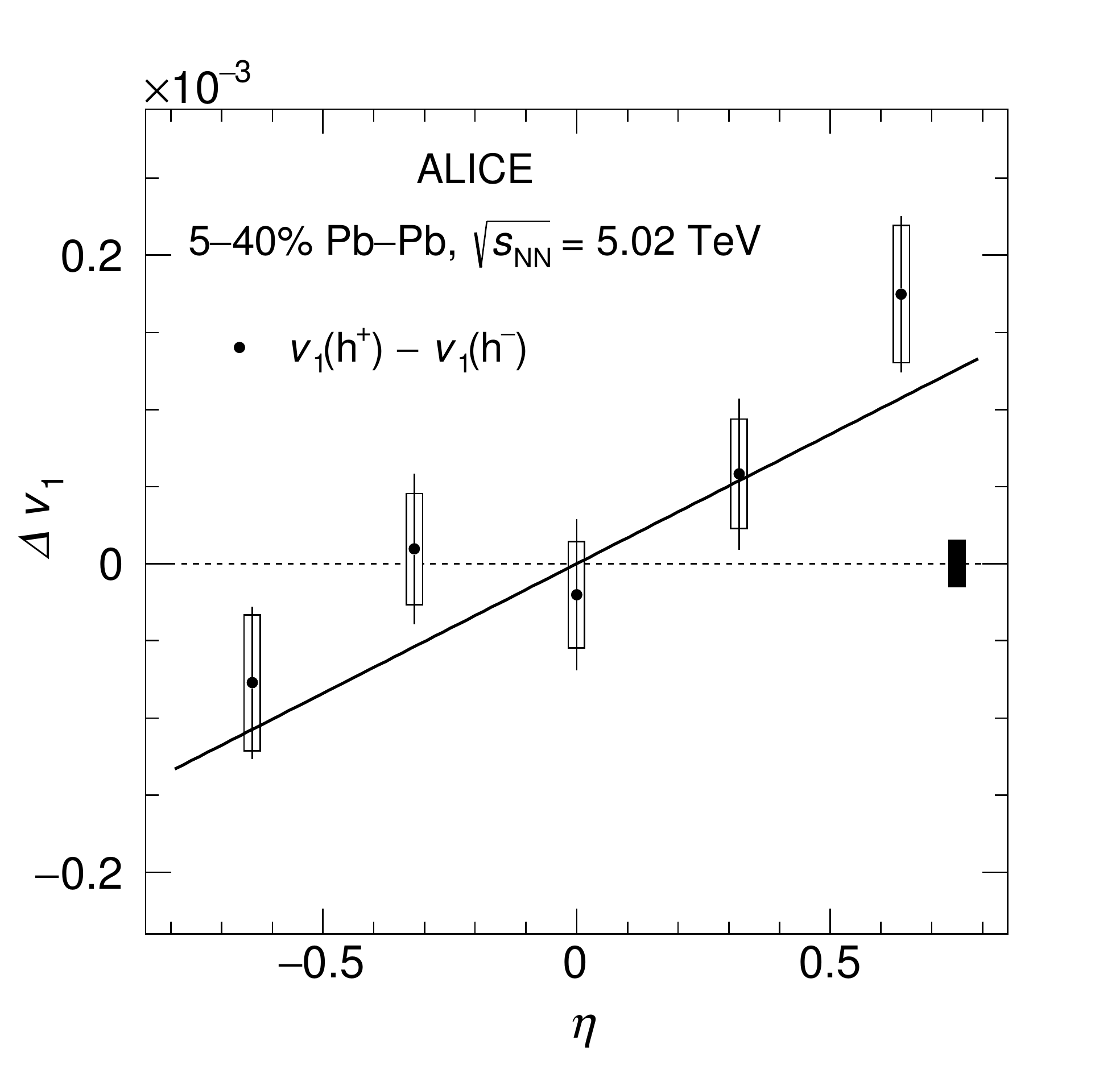}
\includegraphics[scale=0.3]{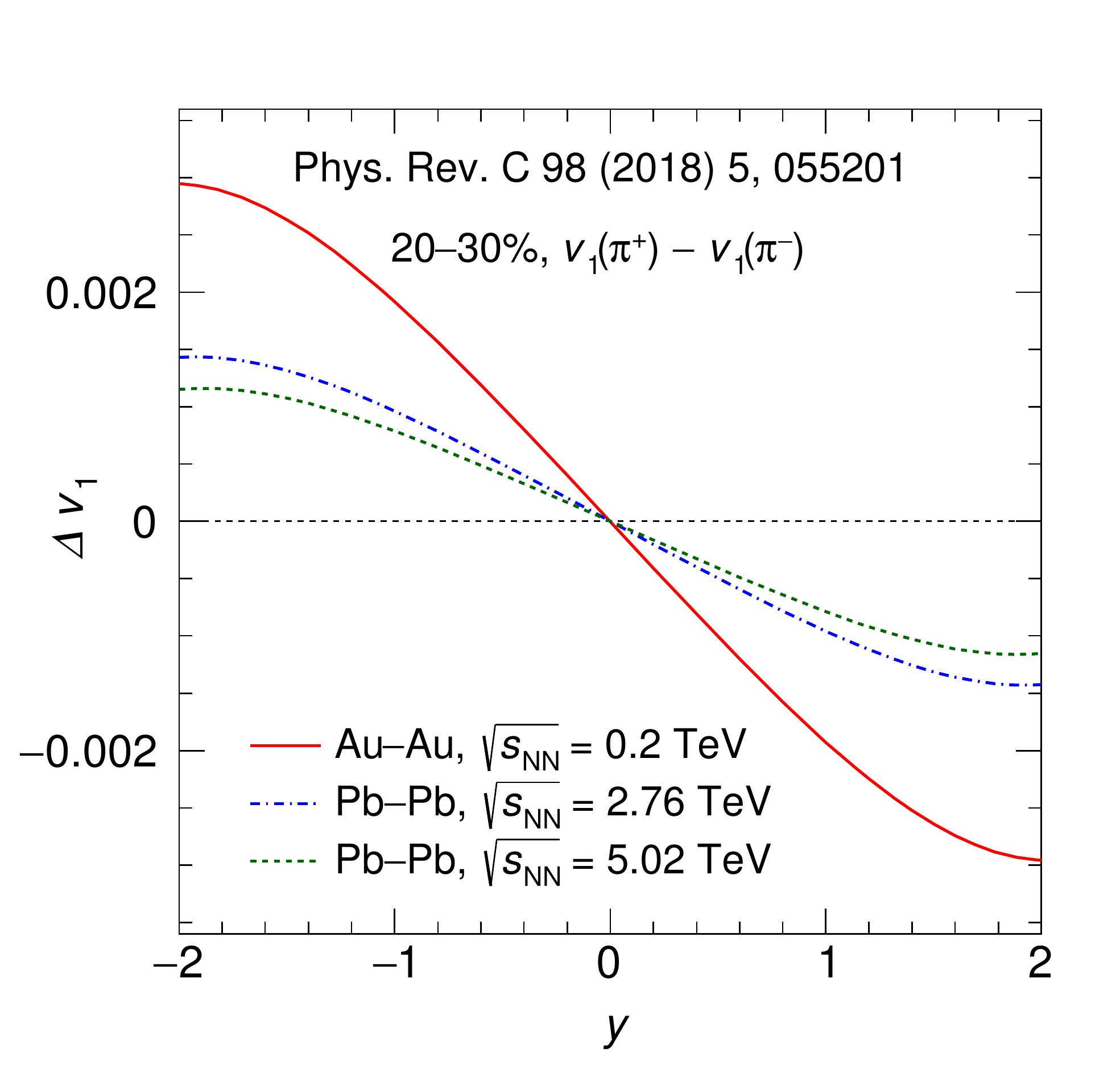}
\caption{ (Left panel) $Delta v_1$ for positively and negatively charged hadrons as a function of pseudorapidity for $p_{\rm T}~>$ 2GeV/$c$ for 5--40\% centrality Pb--Pb collisions at $\sqrt{s_{\rm{NN}}}$ = 5.02 TeV. The solid line represent the fit to the $\Delta v_1$ measurement. (Right panel) Model predictions for the same quantity for pions only in the centrality range 20--30\% at RHIC and LHC energies.}
\label{compare1}
\end{figure}

The ALICE Collaboration measured the pseudorapidity dependence of the directed flow of positively and negatively charged hadrons for the 5--40\% centrality class in Pb--Pb collisions at $\sqrt{s_{\rm NN}}$ = 5.02 TeV~\cite{Acharya:2019ijj}.
The $\Delta{}v_{1}$ for charged hadrons is shown in the left panel of Fig.~\ref{compare1}.

ALICE reported a charge dependent rapidity slope d$\Delta v_1/{\rm d}\eta$, extracted with a linear fit function, of 1.68 $\pm$ 0.49 (stat.) $\pm$ 0.41(syst.) $\times$ $\rm 10^{-4}$ with a significance of 2.6$\sigma$ for being positive.
The model calculations shown in the right panel of Fig.~\ref{compare1} for charged pion $v_1$ yield a similar absolute value of d$\Delta v_1/{\rm d}\eta$ compared to the measured value for charged hadrons, but with an opposite sign.
The comparison at the LHC energies is complicated because the ALICE measurements are only available for unidentified charged hadrons, i.e. the pions, kaons and protons combined.
The physical mechanisms underlying the contributions to the d$\Delta v_1/{\rm d}\eta$ of these different hadrons, could vary significantly among the early-time magnetic field dynamics~\cite{Das:2016cwd, Chatterjee:2018lsx, Chatterjee:2017ahy}, Coulomb interaction with the charged spectators~\cite{Gursoy:2018yai}, and baryon transport to midrapidity via baryon stopping~\cite{Snellings:1999bt}.
While baryon stopping is expected to be less relevant for the LHC energies, its relative effect  on $v_1$ might still be significant, as the $v_1$ signal itself becomes small at those energies. 
Future high precision measurements of charge-dependent directed flow for pions, kaons and protons separately could clarify the role of these various contributions to $v_1$.

\subsection{Heavy-flavor measurements}

The STAR Collaboration reported the first measurement of rapidity-odd directed flow for $\rm{D}^0$ and $\rm\overline{D}{}^0$ mesons at midrapidity ($|y|<$ 0.8) in the 10--80\% centrality interval in Au--Au collisions at $\sqrt{s_{\rm NN}}$ = 200 GeV \cite{Adam:2019wnk}. 
\begin{figure}[ht!]
\centering
\includegraphics[scale=0.3]{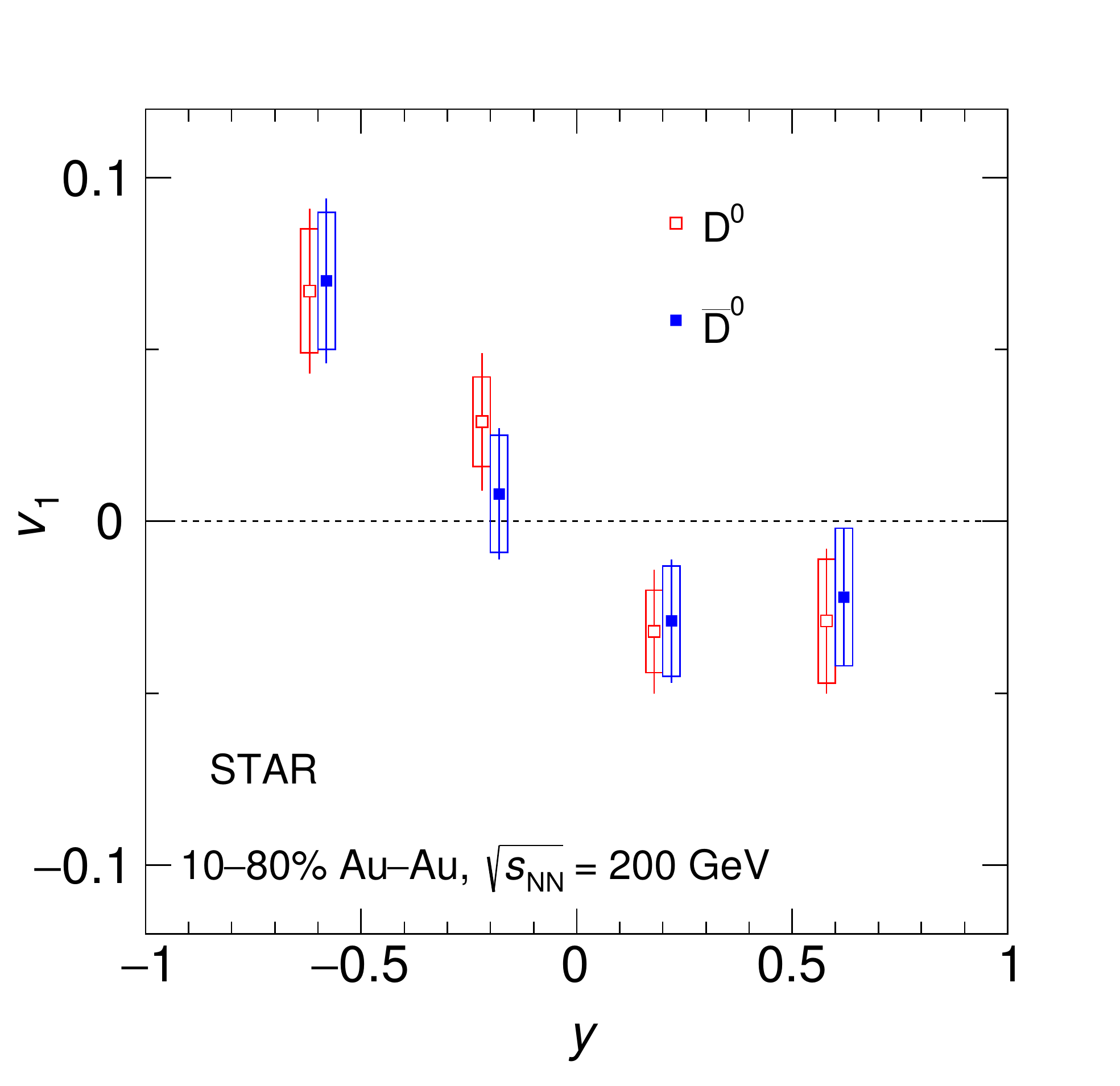}
\includegraphics[scale=0.3]{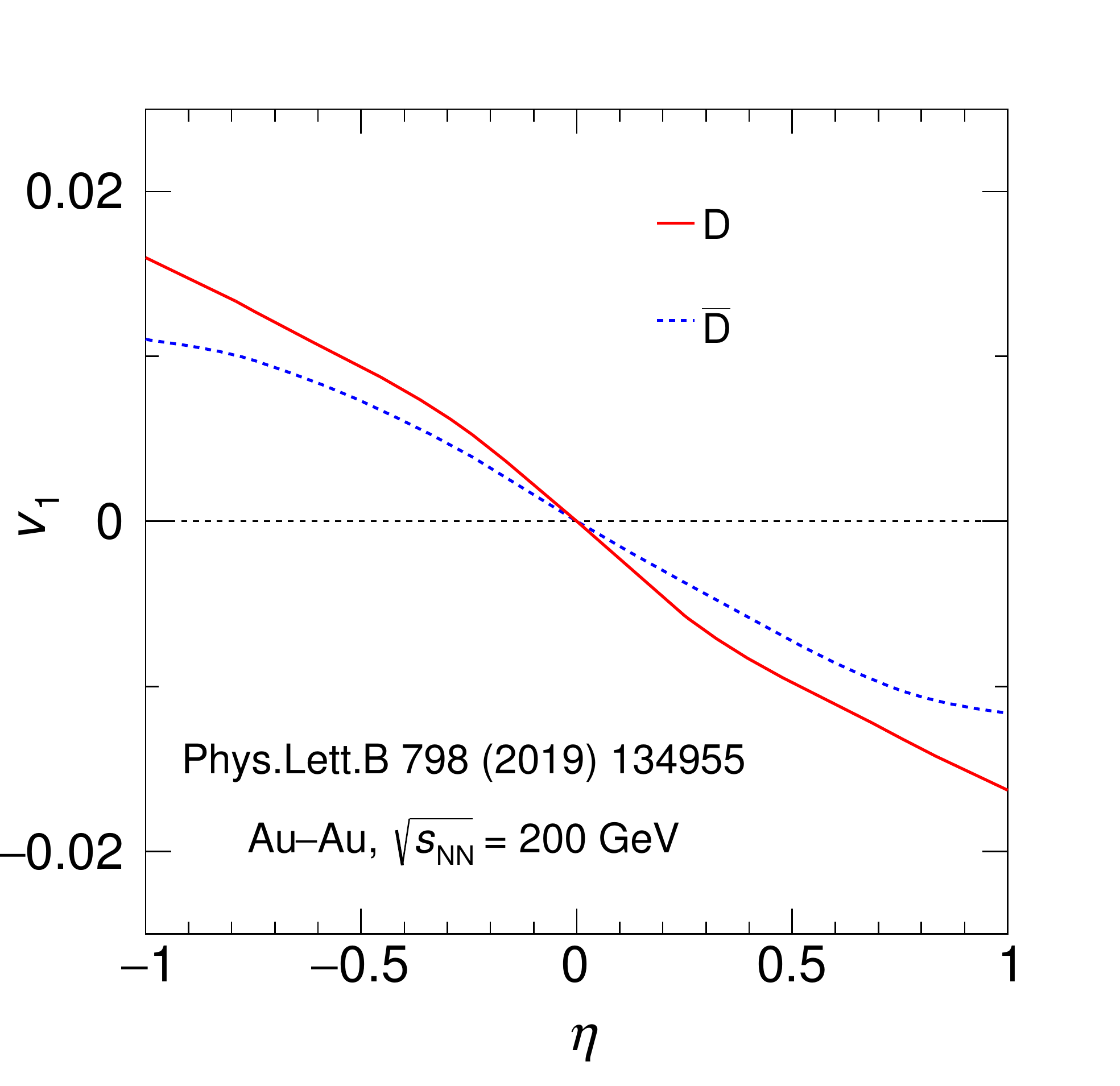}
\caption{ (Left panel) $v_1$ for $\rm{D}^0$ (red markers) and $\rm\overline{D}{}^0$ (blue markers) mesons as a function
of rapidity for $p_{\rm T}~>$ 1.5 GeV/$c$ for 10–80\% centrality Au--Au collisions at $\sqrt{s_{\rm NN}}$ = 200 GeV. The $\rm{D}^0$ and $\rm\overline{D}{}^0$ data points are displaced along the x-axis by for visibility. (Right panel) Rapidity dependence of the directed flow of the D and the $\bar{\rm D}$ mesons from a model calculation which also includes the initial tilt of the fireball in addition to the electromagnetic effects for Au--Au collisions at $\sqrt{s_{\rm NN}}$~=~200~GeV.}
\label{starD}
\end{figure}
The left panel of Fig. \ref{starD} shows the $v_1$ of $\rm{D}^0$
and $\rm\overline{D}{}^0$ for $p_{\rm T}~>$ 1.5 GeV/$c$. The d$v_1$/d$y$ slopes of both $\rm{D}^0$ and $\rm\overline{D}{}^0$ are measured to be negative and are -0.086 $\pm$ 0.025 (stat.) $\pm$ 0.018 (syst.) and -0.075 $\pm$ 0.024 (stat.) $\pm$ 0.020 (syst.), respectively.
The slopes are calculated by fitting $v_1$ as a function of rapidity with a linear function constrained through the origin.
The absolute value of the $\rm{D}^0$-mesons d$v_1$/d$y$ is observed to be about 25 times larger than that of the kaons \cite{Adamczyk:2017nxg} with a 3.4$\sigma$ significance.
The model calculation~\cite{Chatterjee:2018lsx} shown in the right panel of Fig.~\ref{starD} is in qualitative agreement with the STAR measurements. It has been argued that the large d$v_1$/d$y$ for D mesons is mostly driven by the drag from the tilted initial bulk medium. Unlike the bulk itself, the charm quarks are mainly produced by hard binary collisions, which are forward-backward symmetric in rapidity. Consequently, charm quarks produced with non-zero rapidity are shifted with respect to that of light quarks and gluons, resulting in an enhanced dipole asymmetry in the charm quark distribution~\cite{Chatterjee:2017ahy}. As the plasma expands, charm quarks would experience a push from the soft particles in the direction of the shift, leading to larger directed flow of charm hadrons. It is predicted that the contribution to $v_1$ caused by this initial forward-backward asymmetry  dominates over the contribution from the initial EM-field on the D-meson $v_1$~\cite{Chatterjee:2018lsx}.
This calculation, as well as an AMPT model calculation~\cite{PhysRevC.97.064917}, predicts the correct sign of d$v_1$/d$y$ but underestimates the magnitude. However, a realistic non-perturbative heavy quark interaction with the medium, as extracted from the studies of the nuclear modification factor and elliptic flow~\cite{Dong:2019unq}, leads to estimate also quantitatively the strength of d$v_1$/d$y$ for D mesons~\cite{Oliva:2020doe}. Yet, it is hard to draw unequivocal conclusions due to large experimental uncertainties. 

The ALICE Collaboration published the charge dependent directed flow for $\rm{D}^0$ and $\rm\overline{D}{}^0$ mesons as a function of pseudorapidity in the 10--40\% centrality interval in Pb--Pb collisions at $\sqrt{s_{\rm NN}}$ = 5.02 TeV \cite{Acharya:2019ijj}.
The $v_1$ is extracted separately for $\rm{D}^0$ and $\rm\overline{D}{}^0$ mesons, in the $p_{\rm T}$ interval 3--6 GeV/$c$, via a simultaneous fit to the number of D meson candidates and their $v_{1}$ as a function of the invariant mass \cite{Acharya:2019ijj}.
\begin{figure}[ht!]
\centering
\includegraphics[scale=0.3]{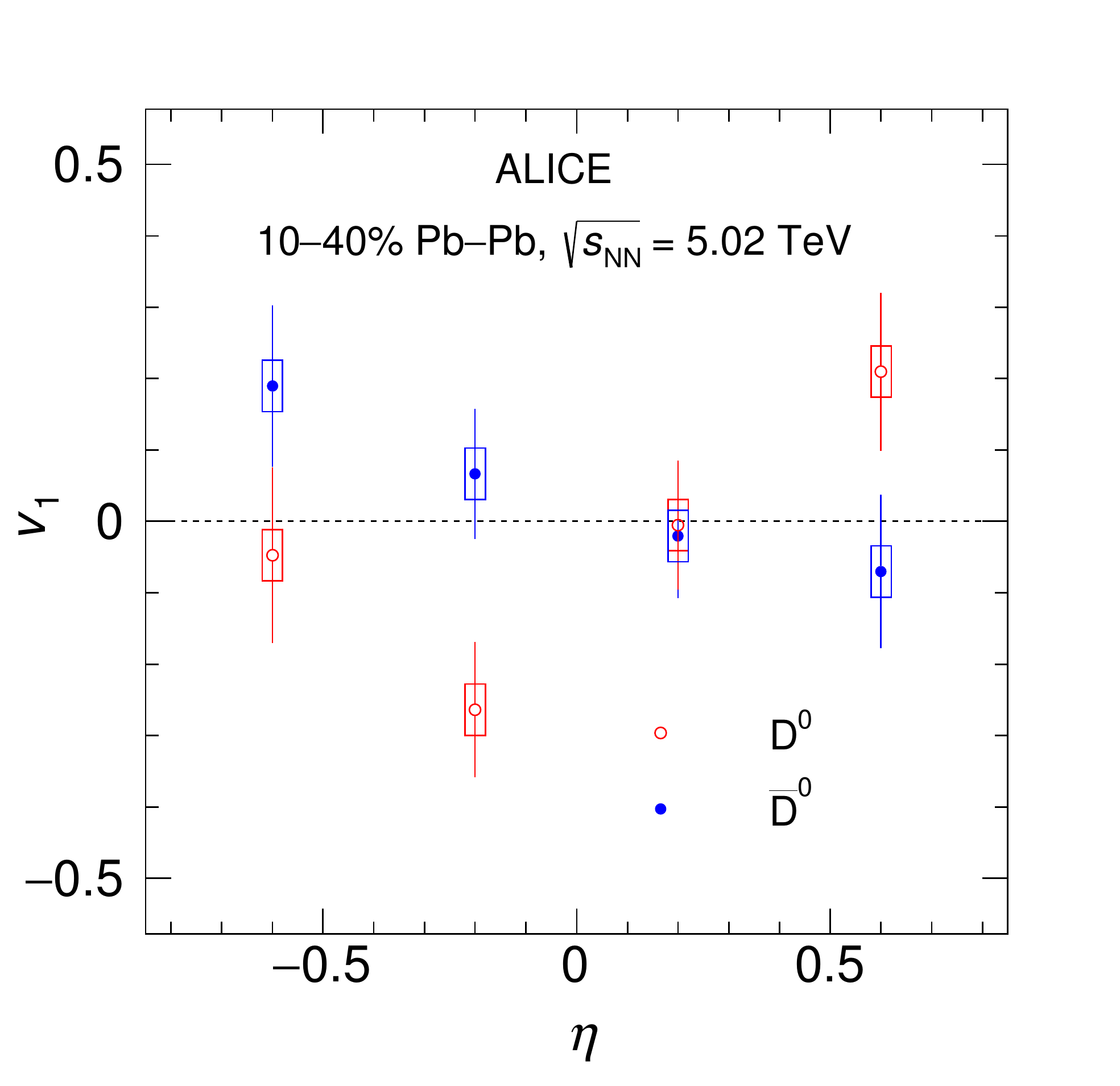}
\includegraphics[scale=0.3]{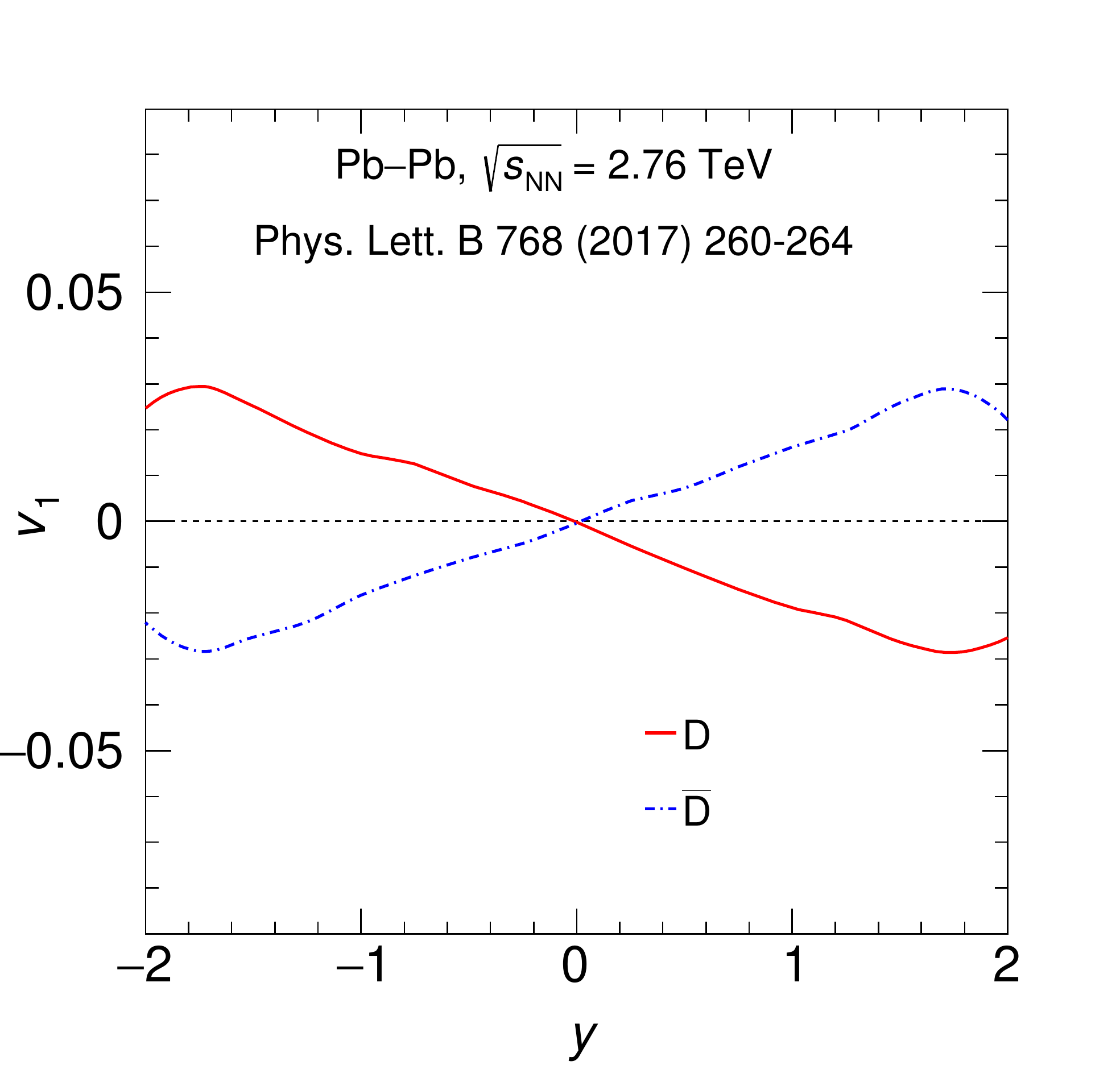}
\caption{
(Left panel) $v_{1}$ of $\rm{D}^0$ (red markers) and $\rm\overline{D}{}^0$ (blue markers) for the 10--40\% centrality interval in Pb--Pb collisions at $\sqrt{s_{\rm NN}}$ = 5.02 TeV. 
(Right panel) Electromagnetically induced directed flow of the $\rm{D}$ (red solid line) and the $\rm{\bar D}$ (blue dash-dotted line) mesons that results from an interplay of  the Lorentz force, the drag and the  stochastic kicks  exerted on the heavy quarks by the bulk of the fireball.}
\label{aliceD}
\end{figure}
The $\rm{D}^0$ and $\rm\overline{D}{}^0$ directed flow as a function of pseudorapidity is shown in the left panel of Fig.~\ref{aliceD}. The measurement indicates a positive slope for the rapidity dependence of the $v_{1}$ of $\rm{D}^0$ and a negative slope for $\rm\overline{D}{}^0$, both with a significance of about 2$\sigma$.
The measurements suggest that an average $v_1$ of $\rm{D}^0$ and $\rm\overline{D}{}^0$ would be consistent with zero. However, the uncertainties on the current experimental results are large and therefore a non-zero charge-integrated $\rm{D}^0$ $v_{1}$ can not be ruled out. 
The $\rm{D}^0$ $v_1$ is an order of magnitude larger and with opposite slope compared to the predictions of the model calculations shown in the right panel of Fig.~\ref{aliceD} and described in Refs.~\refcite{Das:2016cwd}. 
The $v_1$ for $\rm{D}^0$ and $\rm\overline{D}{}^0$ mesons in the 10--40\% centrality interval is about three orders of magnitude larger than the result obtained for charged particles in the 5--40\% centrality class at the same collision energy. 
The $\rm{D}^0$ and $\rm\overline{D}{}^0$ slopes are different from the measurements in Au--Au collisions at $\sqrt{s_{\rm NN}}$~=~200~GeV, where a negative value is observed for both the $\rm{D}^0$ and $\rm\overline{D}{}^0$.

\begin{figure}[ht!]
\centering
\includegraphics[scale=0.3]{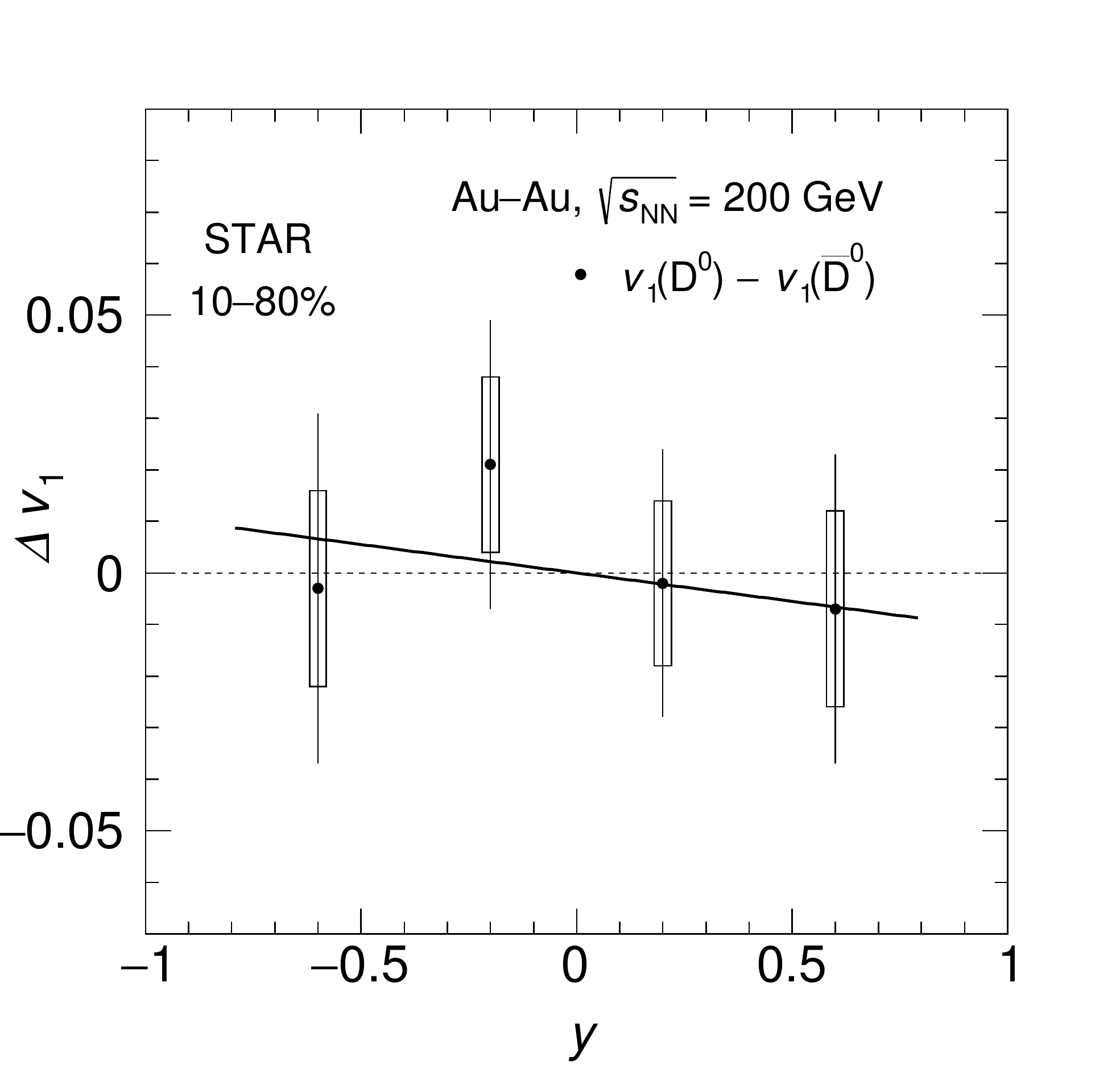}
\includegraphics[scale=0.3]{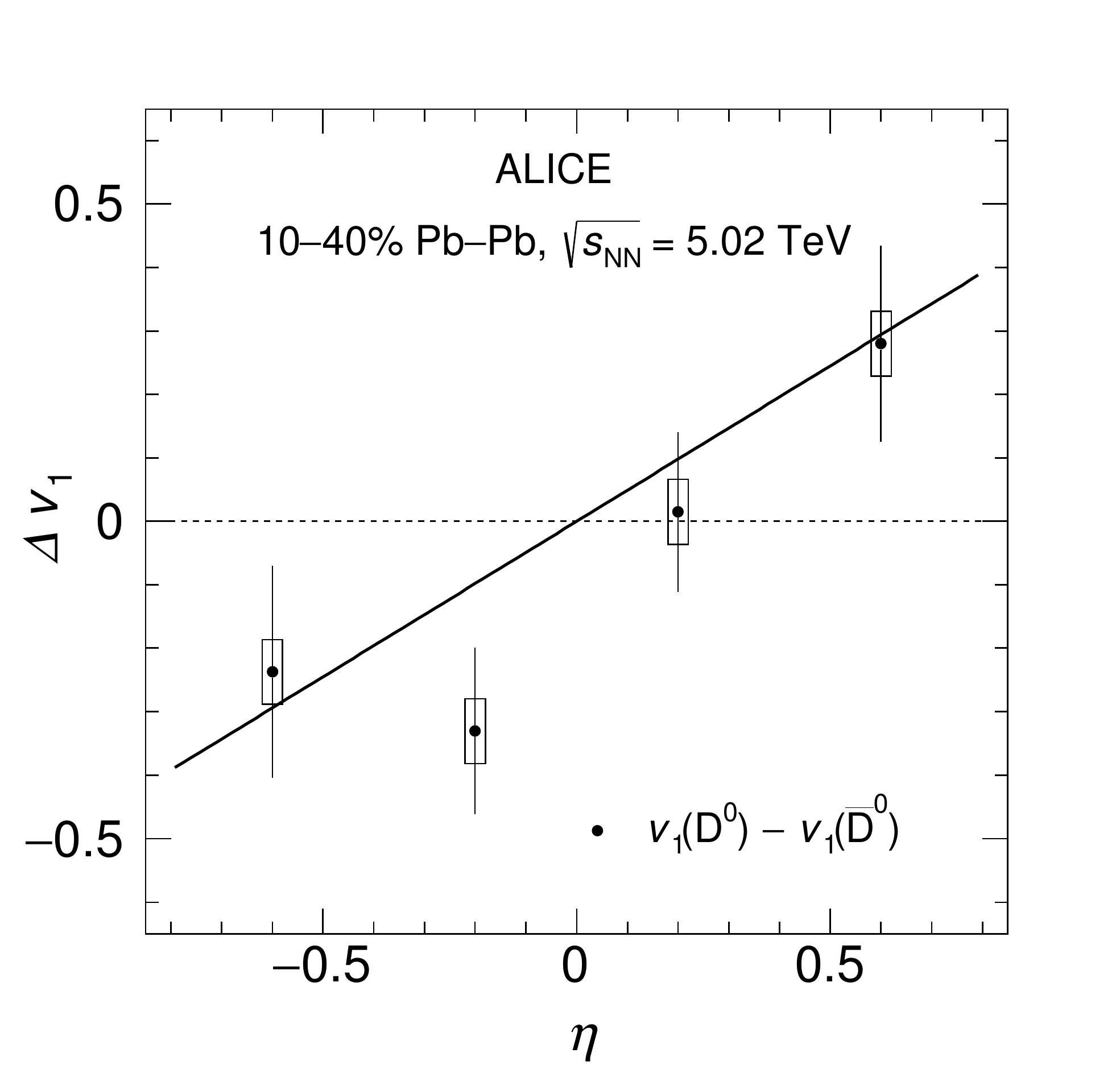}
\vspace{5mm}

\includegraphics[scale=0.3]{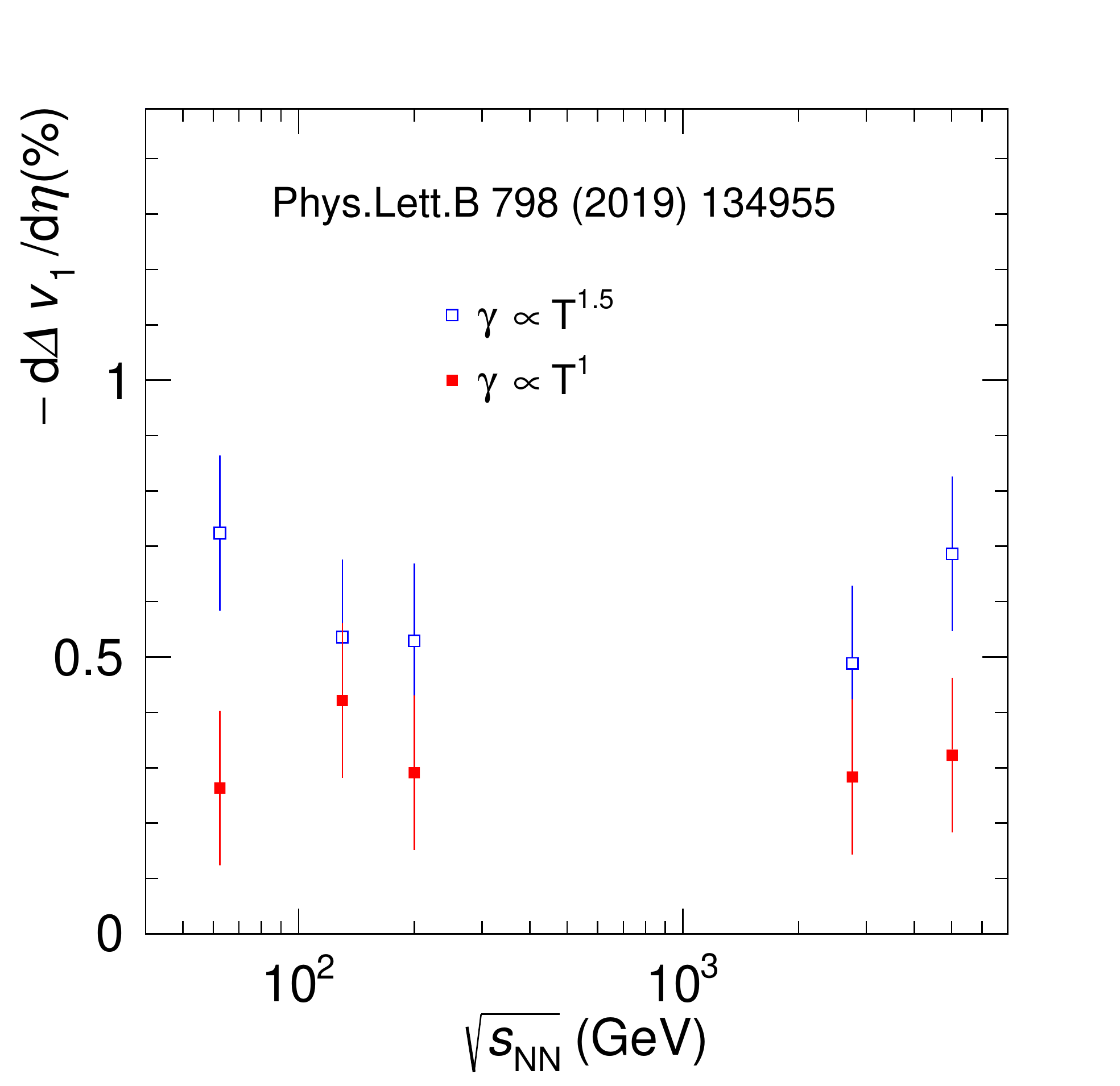}

\caption[]{(Top left panels) $\Delta v1$ between $\rm{D}^0$ and $\rm\overline{D}{}^0$ for $p_{\rm T}~>$~1.5~GeV/$c$ in 10--80\% centrality Au--Au collisions at $\sqrt{s_{\rm NN}}$ = 200 GeV. (Top right panel) $\Delta v_1$ between for $\rm{D}^0$ and $\rm\overline{D}{}^0$ mesons as a function of pseudorapidity for $3~<~p_{\rm T}~>~6$~GeV/$c$ for 5--40\% centrality Pb--Pb collisions at $\sqrt{s_{\rm NN}}$ = 5.02 TeV. The solid lines represent the fit to the $\Delta v1$ measurements.
(Bottom panel) Difference between the mean slopes of the $\rm{D}$ and the $\rm\overline{D}{}^0$ directed flows in rapidity as a function of the collision energy. The calculation also includes the initial tilt of the fireball in addition to the electromagnetic effects (see Ref.~\refcite{Chatterjee:2018lsx}).} 
\label{figDeltav1VSenergy}
\end{figure}
In the top left panel of Fig. \ref{figDeltav1VSenergy} the difference between $\rm{D}^0$ and $\rm\overline{D}{}^0$ directed flow ($\Delta v_1$) measured by the STAR Collaboration in Au--Au collisions at $\sqrt{s_{\rm NN}}$~=~200~GeV is shown.
The $\Delta v_1$ is fitted with a linear function through the origin and it results in a negative slope -0.011 $\pm$ 0.034 (stat.) $\pm$ 0.020 (syst.).
The previously discussed model calculations \cite{Chatterjee:2018lsx,Chatterjee:2017ahy} show a $\Delta v_1$ signal which is smaller than the current precision of the measurement. However, the models predict a $\Delta v_1$ slope for the charm hadrons to be in the range between -0.008 to -0.004. Therefore, small modifications in the values used for the medium conductivity, time evolution of the electromagnetic fields and the description of charm quark dynamics in the model  may result in significant variations in the charge dependent directed flow.

In the top right panel of Fig. \ref{figDeltav1VSenergy} the $\Delta{}v_{1}$ of $\rm{D}^0$ meson measured by the ALICE collaboration at $\sqrt{s_{\rm NN}}$~=~5.02~TeV is shown.
The measured value of d$\Delta v_1/\rm{d}\eta$ = [4.9 $\pm$ 1.7 (stat.) $\pm$ 0.6 (syst.)] $\times 10^{-1}$ corresponds to a significance of 2.7$\sigma$ to have a positive slope. The rapidity slope is once more extracted with a linear fit constrained
to the origin. 
The opposite and large slopes of the $\Delta v_{1}$ of D mesons at the LHC with respect to RHIC energies might indicate a stronger effect of the magnetic field relative to the one due to the induced electric field and the initial tilt of the source, demonstrating sensitivity of the directed flow of charm quark to the interplay among these effects.

Predictions for the dependence of the $\Delta v_1$ for the D meson on the
collision energy is shown in the bottom panel of Fig.~\ref{figDeltav1VSenergy}. The dependence of the charge splitting on collision energy
is predicted to be mostly flat and negative~\cite{Chatterjee:2018lsx}, in contrast to the experimental findings, where an opposite slope between STAR and ALICE is measured.
Dependence of the model results on the choice of the temperature dependence of the drag coefficient is also reported. The model also predicts that the d$v_1$/d$y$ for D mesons can be between 5 and 20 times larger than for charged hadrons depending on the choice of tilt and drag parameters. 
Finally, a more recent model calculation~\cite{Sun:2020wkg} of D mesons found that a $\Delta v_1$ with a magnitude much larger and a sign opposite to the earlier calculations can be generated by assuming a magnetic field with a slower time evolution and with a lifetime of about 0.4 fm/$c$. This brought the model calculations closer to the measurements for heavy flavor.

The CMS Collaboration, following the work published in Ref.~\refcite{Gursoy:2018yai}, measured the elliptic flow of $\rm{D}^0$ and $\rm\overline{D}{}^0$ mesons (2 $<~p_{\rm T}~<$ 8 GeV/$c$) as a function of rapidity for centrality 20--70\% in Pb--Pb collisions at $\sqrt{s_{\rm NN}}$ = 5.02 TeV \cite{CMS-PAS-HIN-19-008}. The difference $\Delta v_2$ between the elliptic flow values of $\rm{D}^0$ and $\rm\overline{D}{}^0$ has been calculated as well. No significant non-zero $\Delta v_2$ is observed within experimental uncertainties. After averaging over the full rapidity range, this results in a value of $\Delta v_2$~=~0.001~$\pm$~0.001~(stat.)~$\pm$~0.003~(syst.). In Ref.~\refcite{Gursoy:2018yai}, the predicted charge-dependent $\Delta v_2$ for charged pions due to the outward electric field (see fig. \ref{figschema}) is negative and with an order of magnitude of about $10^{-3}$ at the LHC energies. Once more, even if not yet significant, the observations possibly indicate a sign opposite to the model.

\section{Future developments}
\label{future}

Measurements from both ALICE and STAR show tantalising signs of electromagnetically induced charge flow in the QGP as predicted by theory. However, currently there is a significant mismatch in the magnitude and sign of the predicted and observed charge-odd flow coefficients. 
The model calculations, reported in Refs.~\refcite{Gursoy:2014aka,Das:2016cwd,Chatterjee:2018lsx}, correctly capture the shape of the $\Delta v_1$ signal for both pions and charm hadrons at RHIC, but predict an opposite sign for the slope in rapidity for protons at RHIC, as well as for both light hadrons and heavy-flavor particles at the LHC. 
This mismatch, however, does not necessarily call for a major revision of the theory. The charge-odd directed flow $\Delta v_1$ results from a very small electric field which remains after cancellation of large opposite sign EM fields. For example, the contributions from the Lorentz and the Faraday current (Fig.~\ref{figschema}) almost always tend to cancel each other with a remnant that is sometimes 2 orders of magnitude smaller. Therefore, various uncertainties in the theoretical values of the transport parameters, or inclusion of heretofore omitted transport channels could easily change the sign and the magnitude of the resulting flow and account for the observed discrepancies.

The theory model described in Sec.~\ref{theory} could be improved in many ways: 
\begin{enumerate}
\item The magneto-hydrodynamic model both for the light and the heavy flavor follows a scheme in which the effects of the electromagnetic fields are appended on the background flow in a perturbative fashion, ignoring the back-reaction of the electromagnetic fields on the background. A fully consistent approach would require solving the hydrodynamics equation coupled to Maxwell's equations in the numerical simulation. A first step in this direction was taken in Ref.~\refcite{Inghirami:2016iru}. 
\item Related to this, new transport channels should be incorporated in the theory. These involve new shear and bulk viscosities~\cite{Dubla:2018czx,Ryu:2015vwa}, which arise from the anisotropy of the background in the presence of electromagnetic fields, Hall conductivity and Hall viscosity~\cite{Amato:2013naa,Feng:2017tsh} and the various forms of anomalous transport~\cite{Kharzeev:2015znc}. 
\item Another example of transport phenomena that was omitted in Sec.~\ref{theory} is {\em shorting}. Just as in an electric circuit, redistribution of charges in a conducting medium is expected to rapidly counterbalance the external electric field hence reduce its magnitude. Clearly this would reduce the electric field that arise from the Coulomb push from the spectators (the Electric fields indicated by the red and blue arrows in Fig. \ref{figschema}) but not the Electric field that arises from the Lorentz effect (which mostly stems from the external magnetic field produced by the spectators instead). Shorting, then, could easily change the balance in favor of the Lorentz electric field, flipping the sign of $\Delta v_1$ and making it consistent with the observation. 
\item There are also systematic and statistical uncertainties in the choice of parameters: the conductivity $\sigma$, the drag and the diffusion coefficients $\mu$ and $D$, and the assumed initial distribution of the nucleons. In particular, the former three depend on the temperature which changes as the plasma expands and cools. This means that these parameters depend on time, a dependence hitherto ignored in the calculations. 
\item Finally, and most interestingly, one should incorporate in the theory, the novel transport phenomena associated to chiral anomalies, namely the Chiral Magnetic and Vortical Effects \cite{Fukushima:2008xe,Kharzeev:2007jp,Tuchin:2010vs,Voronyuk:2011jd,Shi:2017cpu,Kharzeev:2015znc}, and the Chiral Magnetic Wave \cite{Kharzeev:2010gd}. We find it imperative to establish the magnitude and the time profile of the electromagnetic fields, using the more common phenomena discussed in this review, i.e. the electric charge flow induced by the Lorentz and Coulomb forces,  before implementing such exotic effects. 
\end{enumerate} 

On the experimental side, the limited precision of the current measurements prohibits  strong conclusions on the charge transport both for the light- and the heavy-flavor particles.
The following could improve this situation: 
\begin{enumerate}
\item
More precise and differential measurements. These will become possible in the near future with the new data samples that will be collected both at RHIC, in the beam energy scan and fixed target program, and at the LHC, in Run~3~and~4~\cite{Citron:2018lsq, ALICE-PUBLIC-2019-001}. At the LHC this will allow for high precision charge dependent $v_1$ measurements separately for pions and protons and will answer if these have opposite signs. 
\item
Measurement of collisions of isobaric nuclei at RHIC. 
In collisions of $^{96}_{44}$Ru and $^{96}_{40}$Zr one expect to create a very similar system, with very similar flow patterns. At the same time the magnetic field would be proportional to the nuclei charge and can vary by more than 10\%, which can results in a significant modification  in the charge dependent observables. The data have been collected in 2018 by the STAR collaboration and a blind data analysis is under way~\cite{Adam:2019fbq}.
\item
Directed flow measurements of leptons from $Z^0$ decay.
In Ref.~\refcite{Sun:2020wkg} it was found that the role of the slope of particle spectra can strongly affect the sign of the flow. In the specific case of the leptons from $Z^0$ decays one has unique shape of the spectrum, very different from the particle spectra of hadrons and this would induce a new feature in the directed flow measurements that was never discussed before.

\item
New facilities and experiments. In particular the compressed baryonic matter (CBM) experiment~\cite{Golosov:2020brm} at FAIR, the multi-purpose-detector (MPD)~\cite{Parfenov:2019pxf} at NICA and NA61/SHINE~\cite{Kashirin:2020nvw} at CERN SPS will  further explore the phenomena discussed here.

\end{enumerate} 

Establishing an independent probe of the electromagnetic field in heavy-ion collisions is crucial for the search of anomalous transport in the quark-gluon plasma~\cite{Fukushima:2008xe,Kharzeev:2007jp,Tuchin:2010vs,Voronyuk:2011jd,Shi:2017cpu,Kharzeev:2010gd,Kharzeev:2015znc} and for the understanding of the recent experimental observation of global spin polarization of $\Lambda$ baryons~\cite{Becattini:2016gvu,Han:2017hdi,Guo:2019joy}. We argued here that this independent probe may be found in common charge transport. We hope our work will bring new insights and open new avenues for progress in this quest.

\section*{Acknowledgments}
The authors would like to thank Vincenzo Greco and Lucia Oliva for reading the manuscript
and for the valuable suggestions.
A.D. is partially supported by the Netherlands Organisation for Scientific Research (NWO) under the grant 19DRDN011, VI.Veni.192.039. U.G. is partially supported by the Netherlands Organisation for Scientific
26 Research (NWO) under the VIDI grant 680-47-518 and the Delta-Institute for Theoretical Physics (D-ITP), both funded by the Dutch Ministry of Education, Culture and Science (OCW). \\

\bibliographystyle{ws-mpla}
\bibliography{sample}

\end{document}